\documentclass[12pt,3p,doublespacing]{elsarticle}
\usepackage{times}
\usepackage{libertine}
\emergencystretch=\maxdimen
\usepackage{sectsty}
\sectionfont{\color{blue}\large\MakeUppercase}
\subsectionfont{\color{blue}}
\biboptions{authoryear}    
\usepackage{hyperref}
 \hypersetup{
 	plainpages = false, 
	bookmarksopen = true,
	bookmarksnumbered = true,
	breaklinks = true,
	linktocpage,
	colorlinks = true,
	linkcolor = purple,
	urlcolor  = black,
	citecolor = purple,
	}
\usepackage{multicol}
\usepackage{multirow}
\usepackage{imakeidx}
\makeindex
\usepackage{nomencl}
\makenomenclature
\usepackage[xindy]{glossaries}
\makeglossaries
\usepackage{graphics}
\usepackage{graphicx}
\usepackage{epsf,psfrag}
\usepackage{epstopdf}
\usepackage[para]{threeparttable}
\usepackage{subfigure}
\usepackage{epsfig}
\usepackage{pdflscape}
\usepackage{rotating}
\usepackage{lscape}
\usepackage{float}
\usepackage{stfloats}
\usepackage{textgreek}
\usepackage{longtable}
\usepackage{lscape}
\usepackage{capt-of}
\usepackage{tablefootnote}
\usepackage{footnote}
\usepackage{caption}
\usepackage[autopunct=true]{csquotes}
\usepackage{rotating}
\usepackage[normalem]{ulem}
\usepackage{resizegather}
\usepackage{siunitx}
\usepackage[nodisplayskipstretch]{setspace}
\usepackage{color,soul}
\usepackage[usenames,dvipsnames]{xcolor}
\usepackage{amsfonts,amsmath,amssymb,gensymb,mathtools}
\usepackage{latexsym,array}%,subeqn}
\usepackage{textcomp}

\newcommand{\RomanNumeralCaps}[1]
\linenumbers
\makesavenoteenv{tabular}
\makesavenoteenv{table}
\usepackage[left,displaymath, mathlines]{lineno}
\newcommand{\overbar}[1]{\mkern 1.5mu\overline{\mkern-1.5mu#1\mkern-1.5mu}\mkern 1.5mu}
\DeclareMathAlphabet{\mathpzc}{OT1}{pzc}{m}{it}
\def\fig{Fig.~}
\def\figs{Figs.~}
\def\eqn{Eq.~}
\def\eqns{Eqs.~}
\def\tab{Table~}

\def\tsc#1{\csdef{#1}{\textsc{\lowercase{#1}}\xspace}}
\tsc{WGM}
\tsc{QE}
\tsc{EP}
\tsc{PMS}
\tsc{BEC}
\tsc{DE}
\usepackage{easyReview} \setreviewsoff
\newcommand{\removeEq}[1]{\ifistoreview{\@\expandafter\removeColor{#1}\hspace{-0.6em}} \else {}\fi}
\newcommand{\ttxt}[1]{\textbf{\color{blue}\MakeUppercase{#1}}}     
\graphicspath{{../}{Figures/}{Figures/Validation/}}
\DeclareGraphicsExtensions{.png, .PNG,.pdf,.PDF,.jpg,.JPG,.eps,.EPS}
\begin{document}

%\input{reviewer-1.tex}
%\newpage
%\setcounter{page}{1}
%\input{second-review2.tex}
%\newpage
%
%
\setcounter{page}{1}
\begin{frontmatter} % for elsevier jrnls
% 
%\openup -0.5em % 1em for double spacing
\title{\ttxt{Evaluation of RANS-based turbulence models for isothermal flow in a realistic can-type gas turbine combustor application}}
% 
%\title{\ttxt{Electrolyte liquid flow through an asymmetrically charged microfluidic device}}
%
%
\author[labelb]{Aishvarya {Kumar}}
%\ead{jdhakar@ch.iitr.ac.in}
\author[labela]{Ram Prakash {Bharti}\corref{coradd}}\ead{rpbharti@iitr.ac.in}
\address[labela]{Complex Fluid Dynamics and Microfluidics (CFDM) Lab, Department of Chemical Engineering,\\ Indian Institute of Technology Roorkee, Roorkee - 247667, Uttarakhand, INDIA}
\fntext[labelb]{Independent Researcher; Former Researcher, Department of Engineering, City University of London, UK}
%
%\fntext[labelb]{Independent Researcher; Former Researcher, Department of Engineering, City University of London, Northampton Square, London, 4 United Kingdom EC1V 0HB}
%
\cortext[coradd]{\textit{Corresponding author. }}
%
%%%%%%%%%%%%%%%%%%%%%%%%%%%%%%%%%%%%%%%%%%%%%%%%%%%%%%%%%%%%%%%%%%%%%%%%%%%%%%%%%%%%%%
\begin{abstract}
\fontsize{11}{16pt}\selectfont
%----Text of abstract
%
\replace{In the present study, RANS-based turbulence models are assessed to simulate the isothermal flow in a combustor representing a constituent can combustor of the can-annular configuration used in jet engines.  The models assessed are two equation models: standard k-epsilon, realizable k-epsilon, standard k-omega, SST k-omega and Linear Pressure Strain Reynolds Stress Model. The models were assessed by comparing their predictions of mean axial velocity, mean transverse velocity, turbulent kinetic energy and shear stress to experimental data at two different positions in the combustor: the primary holes plane and the dilution holes plane.}{The present study assesses RANS-based turbulence models to simulate the isothermal flow in a combustor representing a constituent can combustor of the can-annular configuration used in jet engines.  The two-equation models (standard $k-\epsilon$, realizable $k-\epsilon$, standard $k-\omega$, SST $k-\omega$), and Linear Pressure Strain - Reynolds Stress Model (LPS-RSM), are assessed by comparing their predictions of mean axial velocity, mean transverse velocity, turbulent kinetic energy, and shear stress with the experimental data at two different positions (i.e., the primary and dilution hole planes) in the combustor.} 
\replace{The comparison showed that the two-equation models failed to predict the confined swirling flow accurately at both positions. The realizable k-epsilon model was the least accurate, followed by the standard k-epsilon model. The standard K-omega performed slightly better, while the SST K-omega model was the most accurate among the two-equation models. The discrepancies between the predicted and experimental results could be attributed to the isotropic turbulence assumptions which are invalid for confined swirling flows and two-equation models also lack formulations to capture the intricacies of vortex flow and its interaction with the surrounding flows in confined swirling flows.}{While the two-equation models generally have failed to predict the confined swirling flow at both positions accurately, the SST $k-\omega$ model yielded the most accurate, followed by standard $k-\omega$ and realizable $k-\epsilon$ models. The discrepancies between the computational and experimental results could be attributed to the isotropic turbulence assumptions, which, however, are invalid for confined swirling flows. Further, the two-equation model formulations cannot capture the intricacies of vortex flow and its interaction with the surroundings in confined swirling flows.}
\replace{The Linear Pressure Strain RSM model, which considers turbulence anisotropy, showed some promise, although overpredicted, results were in trend with experimental values at the primary holes plane. However, at the dilution holes plane, the model overpredicted the velocity field i.e. mean axial velocity and underestimated the turbulence field including turbulent kinetic energy and shear stress. These observed discrepancies can be ascribed to the pressure-strain correlation in the Linear Pressure Strain RMS model which assumes the pressure-strain correlation is a linear function of the strain-rate tensor. However, for complex flows, this linear assumption is too simplistic. Hence, the results of this study suggest that more advanced turbulence models are needed to accurately predict the confined swirling flow in combustors.}{The LPS-RSM, which considers turbulence anisotropy, showed some promise, although overpredicted results follow the trend with experimental values at the primary holes plane. However, at the dilution holes plane, the model overpredicted the velocity field, and underestimated the turbulence field, including turbulent kinetic energy and shear stress. These observed discrepancies can be ascribed to the pressure-strain correlation in the LPS-RSM, which assumes the pressure is a linear function of the strain-rate tensor. However, for complex flows, this linear assumption is quite simplistic. Hence, this study suggests that more advanced turbulence models such as non-LPS-RSM are needed to accurately predict the confined swirling flow in combustors.}
\end{abstract}
%%%%%%%%%%%%%%%%%%%%%%%%%%%%%%%%%%%%%%%%%%%%%%%%%%%%%%%%%%%%%%%%%%%%%%%%%%%%%%%%%%%%%%
%\begin{comment}
%\end{comment}
%%%%%%%%%%%%%%%%%%%%%%%%%%%%%%%%%%%%%%%%%%%%%%%%%%%%%%%%%%%%%%%%%%%%%%%%%%%%%%%%%%%%%%
\end{frontmatter}
%%%%%%%%%%%%%%%%%%%%%%%%%%%%%%%%%%%%%%%%%%%%%%%%%%%%%%%%%%%%%%%%%%%%%%%%%%%%%%%%%%%%%%
%\fontsize{12}{14pt}\selectfont
%\linespread{1.6}
%\doublespacing
%\openup 1em % 1em for double spacing
%\setstretch{2} 
%
%===============================
\section{Introduction}
\label{sec:intro}
%===============================
%
\noindent 
The history of gas turbine engines \replace{goes back to}{began in} the 1920s with {the initial work of} \remove{British Scientist} Sir Frank Whittle\add{, who} first patented the concept in 1930 \citep{Leyes1999}. His design used a compressor to increase the pressure of incoming air, which was mixed with fuel and \remove{then} ignited \replace{ to form}{, resulting in non-premixed (diffusion) combustion, thereby forming} high-velocity gases that drove a turbine\remove{which provided the thrust to move the aircraft}. This rotating turbine \add{provided the thrust to move the aircraft and} also\add{,} in turn\add{,} powered the compressor. 
During the World War II\replace{.}{,} gas turbine \add{(or jet)} engines \remove{(jet engines)} were primarily used in military aircraft\replace{,}{.} \replace{s}{S}ubsequently, their potential for civil aviation was recognized with \add{the} de Havilland Comet becoming the world’s first commercial airliner, powered by four Rolls Royce Avon turbojet engines \citep{Royce2015}.  
Since then, gas turbines have become a prominent power source for commercial and military aircraft as well as for a wide range of applications like power generation and marine propulsion. 
\replace{Advances in technology have led to enhancement in efficiency, lower emissions and higher reliability that have made gas turbines vital for the modern transport and power generation industry. However, with advancements in technology, the principle of operation remains the same for modern gas turbines as those of invented ones.}{Without altering the principle of operation for modern gas turbines, technological advancements have led to enhanced efficiency, lower emissions, and higher reliability that have become vital for the modern transport and power generation industry.}
\\
Gas turbine engines have several primary components \add{(such as compressor, combustor, turbine, and exhaust)} that work together to produce the power to propel aircraft, \replace{generate electricity or}{, and to} provide mechanical \replace{power}{energy}. 
\replace{These components include the compressor which is the first component, it takes air from the atmosphere and compresses it, increasing the air pressure and temperature. The compressed air is then supplied to the combustor. The combustor is where the fuel is injected into the compressed air and ignited, producing high-temperature high pressure which is directed towards the turbine. The turbine is driven by the high-pressure and high-temperature gas from the combustion chamber. The turbine and compressor are assembled on the same shaft, hence, the energy extracted from the gas also drives the compressor and other components like fans (in turbofan jets), generators or even propellers (in the case of turboprop aircraft). The exhaust is the final stage of the gas turbine engine, the hot gas from the turbine is directed out of the engine through the nozzle, producing thrust in the aircraft or when used for power generation the exhaust gas is used in a heat recovery system to capture the remaining heat energy. Thus, the combustor being an essential component of gas turbine engines, its design plays a crucial role in determining the engine’s performance, efficiency and emissions. The combustor is responsible for mixing fuel and igniting the mixture to produce high-temperature and pressure gas. 
}{The combustor is responsible for mixing fuel with the compressed air and igniting the mixture to produce high-temperature, high-pressure gas in the combustion chamber. The energy extracted from the gas drives the turbine and compressor, which are assembled on the same shaft. The exhaust gas is released through the nozzle, producing the thrust in the aircraft or capturing the remaining heat energy for power generation. Thus, the combustor is an essential component of gas turbine engines, and its design plays a crucial role in determining the performance, efficiency, and emissions of the jet engine.}
An ideal\add{,} well-designed combustor ensures complete combustion of the fuel-air mixture, resulting in high combustion efficiency, improved engine performance\add{,}  and reduced emissions. A good combustor must be designed for \remove{high temperatures and pressures, as well as}  harsh operating conditions \add{(e.g., high temperature, high pressure)} of gas turbine engines \remove{. They must be designed} to minimize issues like flameout and flashback \remove{that can occur} during combustion \citep{Lefebvre2010}.  
\\
\replace{There are three main types of combustor geometries used in gas turbine engines, each having its own merits. Can combustor: In this type of system, individual combustion chambers (cans) are located along the engine’s diameter. The can-type combustors were first used in aircraft engines in the 1940s and 1950s \mbox{\citep{Lefebvre2010,Carter1946}}.  The second type of Combustor used in gas turbine engines is Can-annular or tubo-annular. In this system, several combustor cans are arranged in a ring or annulus and cans are connected by interconnecting channels \mbox{\citep{Lefebvre2010}}. Examples include Rolls Royce Spey \mbox{\citep{Gradon1968}}, Rolls Royce Marine Spey \mbox{\citep{Gradon1968a}}, and Pratt and Whitney JT8D \mbox{\citep{Hall1964}}. The third type of combustor chamber design is the annular type. It consists of a ring-shaped combustion chamber that surrounds a turbine shaft. The annular type design allows more complete and efficient combustion, resulting in improved fuel efficiency and lower emissions. Being more compact and having fewer components, the annular design also reduced weight. It is widely used in modern gas turbine engines, particularly suitable for large commercial aircraft examples include General Electric CF6-50 and the Rolls Royce RB211 \mbox{\citep{Lefebvre2010}}.}{Gas turbine engines generally use three types of combustor geometries, each having its own merits, such as the `can combustor' first used in aircraft engines in the 1940s and 1950s \citep{Lefebvre2010,Carter1946}, `can-annular or tubo-annular combustor' \citep{Lefebvre2010} used in Rolls Royce Spey \citep{Gradon1968}, Rolls Royce Marine Spey \citep{Gradon1968a}, and Pratt and Whitney JT8D \citep{Hall1964}, and `annular combustor' widely used in modern gas turbine engines, particularly suitable for large commercial aircraft examples include General Electric CF6-50 and the Rolls Royce RB211 \citep{Lefebvre2010}.}
\replace{The main parts of combustors are the combustor liner which is often made from porous walls, fuel injector, igniter, swirler, primary holes, dilution holes, discharge or exhaust nozzle and combustor liner. The combustor liner is the key component of the combustor, it is part of the combustor that forms the walls of the combustion chamber and contains the hot, high-pressure gases produced by the combustion. The liner must withstand the high temperature and pressures of the combustion process, well as the corrosive effects of the hot gases and possible impurities in the fuel.} {A typical combustor includes the combustor liner (i.e., porous walls of the combustion chamber), fuel injector, igniter, swirler, primary and dilution holes, and discharge (or exhaust) nozzle. The combustor liner is the key component withstanding the possible impurities in the fuel, high temperature/pressure, and the corrosive effects from the gases produced during combustion.}
\replace{The combustor liner geometry can have a significant impact on the performance of the gas turbine engine.}{Thus, the performance of a gas turbine engine significantly depends on combustor liner geometry.} 
The typical combustor liner geometry has \remove{mainly} three main sub-geometries: dome, barrel\add{,} and nozzle.  The \add{primary region of the combustor has} dome, injector, swirler\add{,} and primary holes\remove{ are the parts of the primary region of the combustor}. 
Swirlers \remove{are employed to} create swirling motion in the fuel-air mixture, which \remove{helps} improve\add{s} mixing and achieve\add{s}  complete combustion and lower emissions. 
\replace{Collectively dome geometry of the combustor liner, swirler geometry, fuel injector geometry and placement of the primary holes and their size strongly affect the fuel-air mixing process in the primary region of the combustor thus the overall performance of the combustor making the combustor aerodynamics a highly vital field of study for combustor design. The combustor aerodynamics also assist in the placement of dilution holes that are meant to provide cooler air into the combustor which helps in the even distribution of air-fuel mixture throughout the combustor and even assists in reaching the complete combustion state alongside preventing the development of hot spots that could damage the combustor liner.}
 {Altogether, the geometrical characteristics of the dome, swirler, fuel injection, and the placement (and size) of the primary holes strongly affect the fuel-air mixing, and, thus, the overall performance of the combustor. 
 \\
 Therefore, combustor aerodynamics is a vital field of study for combustor design and assists in the placement of dilution holes, which provides the cooling air to maintain the unfirm distribution of air-fuel mixture and to prevent the development of hot spots which may damage the liner in the combustor.}
The reduction \replace{in the}{of} overall temperature in the combustor also \replace{helps in reducing}{reduces} the production of NO$x$ during the combustion.  
\replace{The exhaust nozzle accelerates the exhaust gas up to high velocity and low pressure and directs it to the turbine \mbox{\citep{Lefebvre2010}}. Hence, the shape and size of the exhaust nozzle of the combustor play a significant role in gas turbine engine performance, and therefore also requires a thorough aerodynamic study to be optimised.}{Further, a thorough aerodynamic study is required to optimize the shape and size of the exhaust nozzle of the combustor, which plays a significant role in gas turbine engine performance, as the nozzle accelerates the exhaust gas up to high velocity and low pressure and directs it to the turbine \citep{Lefebvre2010}.}  
%
%===============================
\section{Literature Review}
\label{sec:intro}
%===============================
%
\replace{Hence, m}{M}any isothermal studies have been performed to enhance the understanding of combustor aerodynamics before combustion flow analysis. \remove{The }LDV (Laser Doppler \replace{Anemometry}{Velocimetry }) technique has often been employed to measure the velocity fields like \replace{M}{m}ean and fluctuating components. Research on realistic models includes stud\replace{ies performed at Imperial College London on}{y of} the combustor can representing the geometry of the Rolls Royce Spey gas turbine engine at different operating conditions \citep{Heitor1986}. 
\replace{The}{For a} combustor \remove{was} made from \remove{a} porous material \replace{T}{t}ransply and \replace{the}{air as} working fluid\remove{was air.}\replace{V}{, v}elocity measurements indicated that the vortical structure in the primary zone is predominantly influenced by \remove{the} swirler flow, with a partial contribution from the primary jets. \remove{The strongly swirling flow was evident from the depression of the profile in the reverse flow region and positive velocities near the axis. The mean swirl velocity profiles demonstrated relatively high values in the primary zone, further amplified in the secondary holes plane due to the effective reduction of the swirling core diameter. The swirl velocity's strength decreased in the exit plane, although the swirl was still present.}
\remove{The }LDV stud\replace{ies have}{y has} also been performed on transparent perspex water combustor \citep{Koutmos1989} \remove{models by the same research group} on similar geometry\add{/conditions \citep{Heitor1986}} \remove{The transparent model provided them with optical access to visualise the development of vortical structures} by injecting dye \replace{in the combustor and also}{and} hydrogen bubbles \add{in the combustor}. \replace{The velocity profiles obtained were in line with those obtained using the Transply combustor, the die displayed the presence of a vortical structure originating due to swirl motion in the primary zone and reaching the combustor exit. }{The velocity profiles complimented excellently in numerical and experimental studies \citep{Heitor1986,Koutmos1989}; the dye displayed a vortical structure originating due to swirl motion in the primary zone and reaching the combustor exit.} The hydrogen bubbles showed the development of a toroidal vortex upstream of the axis of the primary holes. 
\add{\citet{Bicen1989} performed} \replace{Following, a series of experimental studies were conducted on the}{an experimental study using} modified geometry with enlarged \replace{D}{d}ilution \replace{H}{h}oles representing the Rolls Royce Tay combustor geometry \remove{were performed \mbox{\citep{Bicen1989}}}. 
\replace{Through observation, it was noted that jets were formed by water entering the combustor through six primary holes that penetrated almost radially. These jets collided at the centerline, feeding the development of a toroidal recirculation zone in the head of the combustor. The impingement of the primary jets created a strong backflow that transported the swirler flow close to the combustor wall and the primary jets until it reached the dilution plane of dilution jets. Additionally, a small recirculation region was formed near the combustor head (dome) as a result of the swirler flow. Observations also showed that since dilution holes were more staggered because of having a larger diameter, dilution jets blocked the upstream swirler fluid near and wall and forced it to mix with the central core formed by the primary jet impingement Regions of high anisotropy were found at the central core where the tangential turbulence intensity was recorded twice the axial turbulence intensity. Their further research \mbox{\citep{McGuirk1995}} substantiated the above-mentioned.}{They noted that the radial jets, formed by the water entering the combustor through six primary holes, collided at the centerline and developed a toroidal recirculation zone in the head of the combustor. Due to highly staggered dilution holes of larger diameter, dilution jets blocked the upstream swirler fluid near the wall and forced it to mix with the central core, formed by the primary jet impingement regions of high anisotropy, where the tangential turbulence intensity recorded twice of the axial turbulence intensity. These findings were further substantiated \citep{McGuirk1995}.}
\\
\replace{Acknowledging the significance of understanding the aerodynamics of the combustor during the non-reactive stage and the potential of CFD for optimising its design and improving its performance, numerous isothermal CFD studies have been conducted.}{Acknowledging the significance of CFD in understanding aerodynamics, numerous isothermal studies have been conducted to design and improve the performance of the combustor during the non-reactive stage. } Early \remove{numerical simulation} investigations \citep{Srinivasan1980,Sturgess1990,Sloan1986,Nallasamy1987,Xia1998} \remove{performed} on confined swirling flows representing idealized axi\remove{s-}symmetrical can combustors \remove{using the standard k-epsilon model} showed the inadequacy of  standard $k-\epsilon$ model to predict such flows with adverse pressure gradient \remove{accurately} owing to the isotropic eddy viscosity formulation. 
\remove{On realistic geometry, \mbox{\citet{Koutmos1991}} conducted numerical simulations of isothermal flow in a practical single can combustor representing an individual can combustor of the can-annular Rolls Royce Spey engine combustors, using the standard k-epsilon model.  LDV (Laser-Doppler Velocimetry) measurements were used to validate the accuracy of the simulations which showed moderately good agreements at primary and intermediate zones, however, discrepancies increased in the dilution region. Results also showed that the standard k-epsilon model predicted levels of turbulence energy are too low in regions of high anisotropy. Following, \mbox{\citet{McGuirk1993}} evaluated the accuracy of the standard k-epsilon model in predicting the behavior of a practical single can combustor representing an individual can combustor of the tubo-annular Rolls Royce Tay engine combustors. They validated their simulation results using LDV measurements and found that the largest discrepancies between the measured and predicted values were in the primary region of the combustor. The model exhibited higher levels of momentum diffusion than measurement, and the turbulence level was again underpredicted in the primary jet impingement region, which is a highly anisotropic area.}
\add{On realistic geometry, numerical simulations were performed \citep{Koutmos1991,McGuirk1993} to evaluate the accuracy of  the standard $k-\epsilon$ model in predicting the isothermal flow in a practical single can{-annular} combustor  representing an individual can of  can-annular Rolls Royce Spey engine  \citep{Koutmos1991} and the tubo-annular Rolls Royce Tay engine \citep{McGuirk1993}.} 
The Tay combustor differed from the Spey combustor by having larger dilution holes. 
\add{Compared with LDV measurements, the simulation results showed moderately good agreement at primary and intermediate zones; however, discrepancies increased in the dilution region \citep{Koutmos1991} and the largest discrepancies in the primary region of the combustor \citep{McGuirk1993}. The model exhibited higher levels of momentum diffusion and underpredicted turbulence energy in the primary jet impingement, a high anisotropy region.}
\\
\replace{Researchers have attempted to use}{Various studies have used}  the turbulence models with a second-order closure method for simulating confined swirling flows. \add{For instance,} \citet{Jones1989} \remove{performed simulations for swirling confined flows using the standard k-epsilon model and RSM. Comparison with experimental results} showed the superiority of RSM over standard $k-\epsilon$ \add{, in comparison with experimental results, for swirling confined flows,}\replace{.  The researchers noted that the standard k-epsilon model did not incorporate any methodology to consider the stabilizing effects of swirling motion, which resulted in notable differences between the predicted and measured mean velocity fields.}{as the standard k-epsilon model does not incorporate any methodology to consider the stabilizing effects of swirling motion.}  \citet{Weber1990} found \remove{in their investigation} that \add{isotropic turbulent viscosity formulation in} the standard $k-\epsilon$ model\replace{'s isotropic turbulent viscosity formulation was}{, compared with Reynolds Stress model (RSM), is} inadequate in predicting flows with adverse pressure gradients in confined swirling flow applications. \remove{RSM model provided more accurate predictions.} \citet{Lai1996} compared the accuracy of standard k-epsilon, RNG k-epsilon, and RSM models in simulating swirling confined flow\replace{. They}{and} found RSM to be the most accurate \remove{model}, while RNG $k-\epsilon$ was less accurate than standard $k-\epsilon$, particularly at a swirl number of 0.5. \citet{Xia1998} also compared standard $k-\epsilon$, RNG $k-\epsilon$, and \replace{Reynolds Stress models}{RSM} in predicting swirling flow in a water model combustion chamber. \replace{Reynolds Stress}{RSM}  predicted the results most accurately, while the other models \replace{predicted}{displayed} solid-body rotation type flow downstream. \citet{German2005} numerical investigation of the industrial-grade combustor model also showed RSM \remove{model} predicted flow more accurately than the standard $k-\epsilon$ model\replace{. Their prediction also showed}{and} RSM \remove{model} accurately predicted \remove{that} the internal recirculation zone extended to the combustor exit, \replace{while}{but} the standard $k-\epsilon$ model failed\remove{ to do so}. \citet{Jawarneh2006} also showed that RSM \replace{is effective in predicting}{effectively predicts} swirling flow characteristics.
\\
High\replace{ F}{-f}idelity \replace{S}{s}imulations techniques such as \replace{Large Eddy Simulation (LES)}{LES} have also been applied to \replace{model}{explore} isothermal swirling confined flows.
\citet{Jones2012} performed non-reacting \replace{Large Eddy Simulations}{LES} of the idealized model axis\remove{-s}ymmetric  combustor with sudden expansion.  Their \replace{comparison}{results compared reasonably well} with experimental data\replace{showed that they achieved fairly good accuracy and were able to}{, and they could accurately} reproduce the macro features of the confined swirling flows\replace{ accurately}{,} such as \replace{P}{p}recessing \replace{V}{v}ortex \replace{C}{c}ore \add{(PVC)}. \add{Recent studies} \citep{Palkin2022,Wang2022} used LES to simulate isothermal flow in a square combustion chamber, finding good agreement with experimental data but noting a \replace{small}{slight} discrepancy due to insufficient grid resolution in the shear layer. 
\remove{\mbox{\citet{Palkin2022}} performed LES simulation results for isothermal flows in a square combustion chamber that agreed with experimental PIV data, similar to \mbox{\citet{Wang2022}}.} 
However, it can be argued that non\add{e} of the above-mentioned high-fidelity simulation reference cases had essential features of real combustors like primary holes, dilution holes\add{,} and exit nozzle. Hence, the results cannot be extrapolated for realistic geometries. 
\\
The present study focuses on developing sound strategies for \replace{the accurate prediction of}{accurately predicting} swirling flow at non-reacting conditions in a realistic gas turbine combustor at a lower computational cost. This paper strives to enrich the understanding of the research community of modeling swirling flows in gas turbine combustors through accurate prediction and thorough analysis of the flow field \replace{at the}{under} non-reacting conditions. The \remove{authors believe that the} methodology developed in the present research will also help researchers to simulate reacting flow accurately. T\remove{he ultimate aim of t}his study \replace{is}{aims} to provide engineers and designers of various fields from aeronautics to power plants and microturbines with cost-effective strategies for achieving optimal combustor design in less computational time. 
The authors chose the RANS modeling approach due to its efficient and accurate prediction capabilities for the target flows while \remove{also} being relatively less computationally expensive and time-consuming. 
\\
\add{In particular,} this study assesses \remove{ and compares the predictive} the ability of various RANS models like standard $k-\epsilon$ \citep{Launder1972}, realizable $k-\epsilon$ \citep{Shih1995}, standard $k-\omega$  \citep{Wilcox1998}, SST $k-\omega$ \citep{Menter1994} and Linear Pressure Strain Reynolds Stress Model (LPS-RSM) \citep{Gibson1978,Fu1987,Launder1989} to predict the swirl flow in combustor geometry.
\replace{The reference case}{This study has used} \remove{is from} experimental studies \remove{of} \citep{Heitor1986,Heitor1985} under non-reacting conditions \add{ as the reference case}. \replace{The combustor geometry represents an individual can of the tubo-annular Rolls Royce Spey engine combustor. The steady flow conditions were acknowledged before LDV measurements.}{The combustion geometry has been represented as an individual can of the tubo-annular Rolls Royce Spey engine combustor and taken the steady flow conditions from the LDV measurements.} 
To the authors' knowledge, this is the first \add{novel} study \replace{to investigate this in the considered geometry}{considering such a complex combustor geometry}. The \replace{evaluation of the models is performed}{models are evaluated} by comparing \replace{the predicted results}{their predictions} with the experimental Laser Doppler Velocimetry (LDV) data for mean velocity, turbulent kinetic energy\add{,} and shear stress. 
% 
%--------------------------------------
\section{Flow Configuration}
%--------------------------------------
%
\replace{T}{Consider t}he combustor chamber\replace{ is}{, schematically} shown in \fig\ref{fig:1} used in experimental investigations \replace{was}{\citep{Heitor1986,Heitor1985} under non-reacting conditions, as} a model of the tubo-annular or can-annular combustor \replace{representative of}{representing} the Rolls–Royce Spey gas turbine. The combustor model \add{(\fig\ref{fig:1})} include\replace{d}{s} a hemispherical head, a cylindrical barrel, and a circular to rectangular nozzle. 
\begin{figure}[!ht]
	\centering
	\includegraphics[width=1\linewidth]{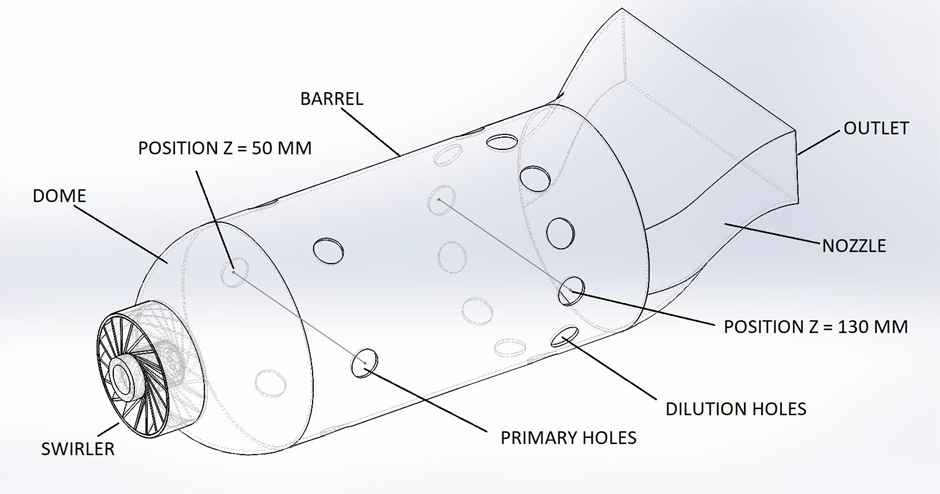}\label{fig:1}
	\caption{Combustor geometry with the positions for velocity measurement.}
	\label{fig:1}
\end{figure}
 \subsection{Reference Case Experimental Setup and Conditions \citep{Heitor1986,Heitor1985}}
 \add{The details of the experimental  setup and conditions are as follows.} Two rows of primary and dilution holes \replace{were}{are} located in the combustor barrel\add{,} with six primary and twelve dilution holes. The \remove{walls of the} combustor \add{walls} \replace{were}{are} constructed using Transply, \remove{which is} a laminated porous material designed by Rolls-Royce plc. The combustor feature\replace{d}{s} a curved vane swirler located on the head, consisting of 18 vanes oriented at 45\degree\ with a maximum thickness of 0.56 mm. The combustor \replace{had}{has} a fuel injector located at the \replace{centre}{center} of the swirler, which \replace{was}{is} in the shape of a 90\degree\ cone with 10 equally spaced 1.7 mm diameter holes located on a 4.50 mm radius through which fuel is injected into the chamber. In this setup, the combustor \replace{was}{is} situated within a sizeable duct, and the primary and dilution jets enter the combustor at an angle of approximately 90\degree\ to its axis.  
 \begin{table}[!b]
 	\centering
 	\caption{Experimental conditions at isothermal run.}\label{tab:1}
 	\scalebox{1}
 	{
 		\begin{tabular}{|c|c|c|c|}
 			\hline
 			Air flow rate (combustor) & Air flow rate (injector nozzle) & $P_0$	& Temperature\\%\hline
 			%	(Kg/s) 	&(g/s) & (atm)	& (K)\\\hline
 			%	0.085	& 1.63 & 1	& 318\\\hline
 			(g/s) 	&(g/s) & (atm)	& (K)\\\hline
 			85	& 1.63 & 1	& 318\\\hline
 		\end{tabular}
 	}
 \end{table}
\\
For isothermal flow conditions, air\add{,} instead of propane\add{,} \replace{was}{is} injected into the combustor through the injector with the same mass flow rate as propane (1.63 g/s)\add{,} implying the average velocity of air through each hole \replace{was}{as} 49 m/s. \add{The bulk flow velocity ($U_b$) is recorded as 31 m/s \citep{Heitor1986,Heitor1985}.} \replace{The atmospheric pressure was 1 atm and the room temperature was around 318K during the experimental studies.}{The experimental \citep{Heitor1986,Heitor1985} conditions are listed in \tab\ref{tab:1}.} 
\replace{The positions on the primary holes z = 50 mm and dilution holes axis z = 130 mm where LDV measurements were taken are}{The LDV measurements are recorded at the positions of primary holes ($z = 50$ mm) and the dilution holes ($z = 130$ mm), as} illustrated in \fig\ref{fig:1}.
\\
\replace{As per the report, of the total mass airflow of 0.085 kg/s, 7\% of the air went into the combustor through the dome section, 14\% of the air went into the combustor through the barrel wall section and 6\% of the air went into the combustor through the nozzle wall section. Therefore, these walls are modelled as mass flow inlet boundary conditions in the present study. Around 24\% of air entered into the injector through a swirler, 16\% of the air went into the injector through primary holes and 33\% of the air went into the injector through dilution holes.}{The total air (flow rate of 0.085 kg/s) enters the combustor \citep{Heitor1986,Heitor1985} through various inlet sections, such as 7\% through the dome, 14\% through the barrel wall, 6\% through the nozzle wall, 24\% through swriler into the injector, 16\% through primary holes into the injector, and 33\% from dilution holes into the injector. The present study models these inlet sections as the mass in-flow boundary conditions (refer \tab\ref{tab:3}).} 
\begin{table}[!bt]
	\centering
	\caption{Boundary Conditions for CFD Simulation - Isothermal Run.}\label{tab:3}
	\scalebox{1}
	{
		\begin{tabular}{|c|c|c|c|c|c|c|}
			\hline
	Boundary	&	Swirler & Primary Holes & Dilution Holes & Dome &Barrel &Nozzle\\\hline
%			(kg/s)&(kg/s) &(kg/s) &(kg/s) &(kg/s) &(kg/s) &(kg/s)\\\hline
%			0.085	& 0.0204	& 0.00227 $\times$ 6 &	0.0023375 $\times$ 12	& 0.00595 &	0.0119 &	0.0051\\\hline
%		&	(g/s)&(g/s) &(g/s) &(g/s) &(g/s) &(g/s) &(g/s)\\\hline
	Air flow (\%)	& 24 & 16 & 33 & 7 & 14 & 6 \\\hline
	Air flow (g/s)	& 20.4	& 2.2667 $\times$ 6 &	2.3375 $\times$ 12	& 5.95	 &	11.9 &	5.1	\\\hline
		\end{tabular}
	}
\end{table}
\remove{The experimental conditions are listed in \mbox{\tab\ref{tab:1}} and boundary conditions are listed in \mbox{\tab\ref{tab:3}}. The combustor geometry and experimental details are described in greater detail in \mbox{\citep{Heitor1986,Heitor1985}}. A schematic representation of the combustor is shown in \mbox{\fig\ref{fig:1}}.}
%
%--------------------------------------
\section{Methodology}
%--------------------------------------
%
\add{The mathematical model for the problem under consideration is written as follows.}
\subsection{Governing Equations}
Steady\add{-}state incompressible RANS (Reynolds-Averaged Navier-Stokes) equations  \citep{Versteeg2007,Alfonsi2009} have been employed and have the following form: 
\begin{gather}
	\frac{\partial{\overbar{u}}_i}{\partial x_i}=\ 0\ 
	\label{eq:1} \\
	\rho\frac{\partial{\overbar{u}}_i}{\partial t}+\ \rho{\overbar{u}}_j\frac{\partial{\overbar{u}}_i}{\partial x_j}=\rho{\overbar{f}}_i+\ \frac{\partial\ }{\partial x_i}\left[-\overbar{p}\delta_{ij}+2\mu{\overbar{S}}_{ij}\ -\rho\overbar{u_i^\prime u_j^\prime}\right]
	\label{eq:2}
	\\
	{\overbar{S}}_{ij}=\ \frac{1}{2}\ \left[\frac{\partial{\overbar{u}}_i}{\partial x_j}+\ \frac{\partial{\overbar{u}}_j}{\partial x_i}\right]
	\label{eq:3}
\end{gather}
Where ${\overbar{S}}_{ij}$ is the mean rate of the strain tensor, ${\overbar{u}}_i$ is the mean fluid velocity in $i^{th}$ direction, $\rho$ is the density of the fluid,  $\overbar{p}$ is the mean pressure, $\mu$ is the dynamics viscosity, ${\overbar{f}}_i$ is the source term for body forces like gravity, $\delta_{ij}$ is Kronecker delta\remove{and is 1 when $i = j$  and 0 when $i\ \neq j$},  and ($-\rho\overbar{u_i^\prime u_j^\prime}$) is the apparent stress owing to the fluctuating velocity field and is generally referred to as Reynolds stress. 
\\
In this study, the Reynolds stress \add{($-\rho\overbar{u_i^\prime u_j^\prime}$)} for two-equation models is modelled using Boussinesq's approximation, which is defined below.
\removeEq{
\begin{gather*}
-\rho\overbar{u_i^\prime u_j^\prime}=\ {2\mu}_t\left[\frac{1}{2}\left(\frac{\partial{\overbar{u}}_i}{\partial x_j}+\ \frac{\partial{\overbar{u}}_j}{\partial x_i}\right)\right]-\ \frac{2}{3}{\rho k\delta}_{ij}
	\label{eq:4} 
\end{gather*}
}
\remove{which can be shorthanded as}
\begin{gather}
	-\rho\overbar{u_i^\prime u_j^\prime}=\ {2\mu}_t\left[{\overbar{S}}_{ij}\right]-\ \frac{2}{3}\rho{k\delta}_{ij}
	\label{eq:5}
\end{gather}
Where $k$ \remove{is} the turbulent kinetic energy \add{is} defined \replace{by the following equation}{as follows.}
\begin{gather}
	k=\ \frac{1}{2} \left[\left(\overbar{u_i^\prime}\right)^2 + \left(\overbar{u_j^\prime}\right)^2 +\left(\overbar{u_k^\prime}\right)^2 \right]
	\label{eq:6}
\end{gather}
\replace{As above mentioned, the constant temperature was maintained in the test rig in the experimental study, therefore, the process is considered isothermal and hence energy equation is not used. The derivations of the above equations can be found in \mbox{\citep{Versteeg2007,Alfonsi2009}}.}{As mentioned {above}, the experimental data is available at constant temperature; thus, the process is considered isothermal, and the energy equation is not required.}
%
%--------------------------------------
\subsection{Turbulence Models}
%--------------------------------------
\add{The turbulence models, such as standard $k-\epsilon$, realizable $k-\epsilon$, standard $k-\omega$, SST $k-\omega$, and Linear Pressure Strain - Reynolds Stress Model (LPS-RSM), used in this study are expressed as follows.}
\subsubsection{Standard $k-\epsilon$ model}\label{sec:ske}
In the standard $\kappa-\epsilon$ model \citep{Launder1972}, the transport equation\add{s} of turbulent kinetic energy ($k$), and turbulent dissipation rate ($\epsilon$) \remove{is obtained from the following equations:} \add{are as follows.} 
\begin{gather}
\frac{\partial\rho k}{\partial t}+\ \frac{\partial}{\partial x_j}\left(\rho k{\overbar{u}}_j\right)=\ \frac{\partial}{\partial x_j}\left[\left(\mu+\ \frac{\mu_t}{\sigma_k}\right)\frac{\partial k}{\partial x_j}\right]+2\mu_t{\overbar{S}}_{ij}\cdot{\overbar{S}}_{ij}-\rho\epsilon	
\label{eq:7}\\
	\frac{\partial\rho\epsilon}{\partial t}+\ \frac{\partial}{\partial x_j}\left(\rho\epsilon{\overbar{u}}_j\right)=\ \frac{\partial}{\partial x_j}\left[\left(\mu+\ \frac{\mu_t}{\sigma_\epsilon}\right)\frac{\partial\epsilon}{\partial x_j}\right]+C_{1\epsilon}\frac{\epsilon}{k}2\mu_t{\overbar{S}}_{ij}\cdot{\overbar{S}}_{ij}-C_{2\epsilon}\frac{\epsilon^2}{k}+S_\epsilon
\label{eq:9}
\end{gather}
where \remove{$k$ is the turbulent kinetic energy,}  $\mu_t$ is the turbulent (or eddy) viscosity \add{assumed to be isotropic}\replace{,}{and} $\epsilon$ is the turbulence dissipation rate \remove{, and is defined by the following term} \add{are defined as follows}.
\begin{gather}
\epsilon=2\nu\overbar{s_{ij}^\prime s_{ij}^\prime}
	\label{eq:8} \\
	\mu_t= \rho C_\mu\frac{k^2}{\epsilon}
	\label{eq:10} 
\end{gather}
\remove{where $C_\mu$ is a constant.} The model constants ($C_{1\epsilon}$,  $C_{2\epsilon}$,  $C_\mu$, $\sigma_k$, $\sigma_\epsilon$) have the following values.
\begin{gather*}
	C_{1\epsilon}=1.44, \qquad C_{2\epsilon}=1.92, \qquad  C_\mu=0.09, \qquad  \sigma_k=1.0,  \qquad \sigma_\epsilon=1.3
	\label{eq:11} 
\end{gather*}
These values have been experimentally determined \citep{Launder1972} for fundamental turbulent flows\add{,} including frequently encountered shear flows like boundary layers, mixing layers, and jets\add{,} as well as for decaying isotropic grid turbulence. 
\subsubsection{Realizable $k-\epsilon$ model} \label{sec:rke}
\add{In the realizable $k-\epsilon$ model, the transport equation for turbulence kinetic energy ($k$) remains same (i.e., \eqn\ref{eq:7}) as the standard $k-\epsilon$ model; however, the transport equation of turbulent dissipation rate ($\epsilon$) is derived using the transport equation of mean square of vorticity fluctuation \citep{Tennekes2018}.} 
The realizable $k-\epsilon$ model \citep{Shih1995}\remove{, the turbulent kinetic energy ($k$) and turbulent dissipation rate ($\epsilon$) are obtained from the following} equations \add{are written as follows}: 
\begin{gather}
\frac{\partial\rho k}{\partial t}+\ \frac{\partial}{\partial x_j}\left(\rho k{\overbar{u}}_j\right)=\ \frac{\partial}{\partial x_j}\left[\left(\mu+\ \frac{\mu_t}{\sigma_k}\right)\frac{\partial k}{\partial x_j}\right]+2\mu_t{\overbar{S}}_{ij}\cdot{\overbar{S}}_{ij}-\rho\epsilon	
\label{eq:12}  \\
	\frac{\partial\rho\epsilon}{\partial t}+\ \frac{\partial}{\partial x_j}\left(\rho\epsilon{\overbar{u}}_j\right)=\ \frac{\partial}{\partial x_j}\left[\left(\mu+\ \frac{\mu_t}{\sigma_\epsilon}\right)\frac{\partial\epsilon}{\partial x_j}\right]+\rho C_{1}{S}\epsilon-\rho C_{2}\frac{\epsilon^2}{k+\sqrt{\nu\epsilon}}
\label{eq:15}\\
%\end{gather}
%\remove{The following is the transport equation of $\epsilon$}
%\begin{gather}
	\text{where}\qquad C_1=\max\left(0.43, \frac{\eta}{\eta+5}\right),
	\qquad \eta=S\left(\frac{k}{\epsilon}\right),\qquad  S=\ \sqrt{2\ {\overbar{S}}_{ij}\cdot{\overbar{S}}_{ij}} \nonumber\label{eq:16} \\
	C_2=1.9,\qquad C_{1\epsilon }= 1.44,\qquad \sigma_k=1,\qquad {\ \sigma}_\epsilon=1.2
	\nonumber\label{eq:17} 
\end{gather}
where\add{,} \remove{$k$ is the turbulent kinetic energy, }  $\epsilon$  \remove{is the turbulence dissipation rate and,} \add{and} $\mu_t$ \remove{is the turbulent viscosity or eddy viscosity} \replace{and is}{are} defined \remove{by the following term which is} similar\add{ly} to the standard $k-\epsilon$ model \add{by \eqns(\ref{eq:8}) and (\ref{eq:10}), respectively}. 
\removeEq{
\begin{gather*}
	{\epsilon}=2\nu\overbar{s_{ij}^\prime s_{ij}^\prime}
	\label{eq:13}  \\ 
	\mu_t= \rho C_\mu\frac{k^2}{\epsilon}
	\label{eq:14}    
\end{gather*}
}
\remove{The eddy viscosity in the realizable $k-\epsilon$ model is also assumed to be isotropic and is computed using the following equation similar to the standard $k-\epsilon$ model: }
\removeEq{
	\begin{gather*}
	\mu_t= \rho C_\mu\frac{k^2}{\epsilon}
	\label{eq:14} 
\end{gather*}
}
To be realizable, the model requires maintaining non-negativity of normal stress ($\overbar{{{u^\prime}_i}^2}\geq0$) and ``Schwarz inequality" \citep{Heitor1985,Launder1972}\replace{. The Schwarz inequality is defined by the following term}{expressed as follows}.
\begin{gather}
	\overbar{{{u^\prime}_i}^2}\geq0
	\qquad\text{and}\qquad
	{\frac{(\overbar{u_i^\prime\ u_j^\prime})}{\overbar{u_i^{\prime2}}\ \overbar{u_j^{\prime2}}}}^2\le1
	\label{eq:18} 
\end{gather}
\add{Schwarz inequality binds the magnitude of shear stress in a fluid and ensures that the stresses are realistic and physically meaningful. The turbulence velocity formulation remains isotropic. The realizability of the model is achieved by modifying $C_\mu$ in \eqn(\ref{eq:10}), which is constant in the standard $k-\epsilon$ model, by introducing the additional terms and coefficients that depend on the strain and rotation rates of the flows and aimed to assist in improving the turbulence-predicting ability, as follows.}
%The realizability is achieved by formulating $C_\mu$ in \eqn(\ref{eq:10}) by the following equation.
%
\begin{gather}
	C_\mu= \frac{1}{A_0+A_S{({kU^*}/{\epsilon})}}
	\label{eq:19} \\
	\text{here}\qquad U^\ast\equiv\sqrt{{\overbar{S}}_{ij}\cdot{\overbar{S}}_{ij}+{\widetilde{\Omega}}_{ij}{\widetilde{\Omega}}_{ij}}
	\label{eq:20} \\
	\text{and}\qquad{\widetilde{\Omega}}_{ij}=\ {\overbar{\Omega}}_{ij}-\nabla\times\omega_k,
	\qquad 
	{\overbar{\Omega}}_{ij} = \frac{1}{2}\left[\frac{\partial{\overbar{u}}_i}{\partial{\overbar{x}}_j}-\frac{\partial{\overbar{u}}_j}{\partial{\overbar{x}}_i}\ \right]
	\label{eq:21} 
\end{gather}
where ${\overbar{\Omega}}_{ij}$ is the mean rate of rotation tensor, viewed in a moving reference frame with angular velocity ($\omega_k$). The model constant $A_0$ and $A_s$ are given as 
\begin{gather}
	A_0=\ 4.04,\quad  A_s=\sqrt6cos\phi,\quad  \phi=\ \frac{1}{3}\cos^{-1}{(\sqrt6}W),\quad  W=\ \frac{{\overbar{S}}_{ij}\ {\overbar{S}}_{jk}{\overbar{S}}_{ki}}{{\widetilde{S}}^3},\quad  \widetilde{S}=\ \sqrt{{\overbar{S}}_{ij}{\overbar{S}}_{ij}}
	\label{eq:22} \nonumber
\end{gather}
These terms are aimed to assist in improving the turbulence-predicting ability of the model. 
\subsubsection{Standard $k-\omega$ model}
The turbulent kinetic energy \replace{,}{(}$k$\replace{,}{)} and specific dissipation rate \replace{,}{(}$\omega=\epsilon/k$\replace{, which is a ratio of $\epsilon$ to $k$,}{)} are obtained from the following equation in the standard $k-\omega$ model \citep{Wilcox1998}. 
\begin{gather}
\frac{\partial\rho k}{\partial t}+\ \frac{\partial}{\partial x_j}\left(\rho k{\overbar{u}}_j\right)=\ \frac{\partial}{\partial x_j}\left[\left(\mu+\ \frac{\mu_t}{\sigma_k}\right)\frac{\partial k}{\partial x_j}\right]+2\mu_t{\overbar{S}}_{ij}\cdot{\overbar{S}}_{ij}-\rho\beta^*f_{\beta^*}k\omega	
	\label{eq:23} \\
\frac{\partial\rho \omega}{\partial t}+\ \frac{\partial}{\partial x_j}\left(\rho \omega{\overbar{u}}_j\right)=\ \frac{\partial}{\partial x_j}\left[\left(\mu+\ \frac{\mu_t}{\sigma_\omega}\right)\frac{\partial \omega}{\partial x_j}\right]+\alpha\left(\frac{\omega}{k}\right)2\mu_t{\overbar{S}}_{ij}\cdot{\overbar{S}}_{ij}-\rho\beta f_{\beta}\omega^2	
\label{eq:24} 
\end{gather}
\add{where, $\sigma_k=2.0$, and $\sigma_\omega=2.0$.} The isotropic eddy viscosity \add{($\mu_t$)} is computed \replace{from the following equation:}{by}
\begin{gather}
\mu_t=\alpha^\ast\frac{\rho k}{\omega}
	\label{eq:25} 
\end{gather}
%
%\subsubsection*
\textit{(a) Low Reynolds number correction: }
The coefficient $\alpha^\ast$ \add{(in \eqn\ref{eq:25})} damps the turbulent viscosity causing a low Reynolds number ($Re$) correction given by: 
\begin{gather}
	\alpha^\ast=\ \alpha_\infty^\ast\left(\frac{\alpha_0^\ast+Re_k}{1+ Re_k}\right)
	\label{eq:26}  \\
	\text{where}\qquad 
	Re_k=\ \frac{Re_t}{R_k},\qquad
	Re_t=\ \frac{\rho k}{\mu_t\omega},\qquad
	R_k=6, \qquad \alpha_o^\ast=\ \frac{\beta_i}{3}, \qquad
	\beta_i=0.072, 	\qquad \alpha_\infty^\ast = 1
	\label{eq:27} \nonumber
\end{gather}
In the high Reynolds number form of the $k-\omega$ model, $\alpha^\ast=\ \alpha_\infty^\ast=1$. 
%
%\subsubsection*
\\
\textit{(b) Production of ${\omega}$ \label{sec3.2.3.2}: }
\replace{In the production term $\alpha({\omega}/{k})2\mu_t{\overbar{S}}_{ij}\cdot{\overbar{S}}_{ij}$ in \mbox{\eqn(\ref{eq:24})}, t}{T}he coefficient $\alpha$ \add{in the production term ($\alpha({\omega}/{k})2\mu_t{\overbar{S}}_{ij}\cdot{\overbar{S}}_{ij}$) in \eqn(\ref{eq:24}),} is given by
\begin{gather}
	\alpha = \ \frac{\alpha_\infty}{\alpha^\ast}\left(\frac{\alpha_0+Re_\omega}{1+Re_\omega}\right),
	\quad Re_\omega=\frac{Re_t}{R_\omega},\quad R_\omega=2.95, \quad \alpha_\infty=0.52, \quad \alpha_0= \frac{1}{9}
	\label{eq:28} 
\end{gather}
where \remove{$Re_\omega=({Re_t}/{R_\omega})$, $R_\omega=2.95$,} $\alpha^\ast$ and ${Re}_t$ are given by \eqn(\ref{eq:26}). 
\remove{In the high Reynolds number form of the $k-\omega$ model, $\alpha^\ast=\ \alpha_\infty^\ast=1$.}
%
%\subsubsection*
\\
\textit{(c) Modelling turbulence dissipation: }
\add{The factors ($f_{\beta^\ast}$, $\beta^\ast$,  $f_{\beta}$ and $\beta$)} in the dissipation term\add{s} ($\rho\beta^\ast f_{\beta^\ast}k\omega$\add{, and $\rho\beta f_\beta\omega^2$}) \remove{of the transport equation of $k$} in \eqns(\ref{eq:23}) \add{and (\ref{eq:24}) are given by} 
\begin{gather}
f_{\beta^\ast}=
\begin{cases}
	1 &\chi_k\le0\\
	{(1+680\chi_k^2)}/{(1+400\chi_k^2)} &\chi_k>0
\end{cases},
\qquad\qquad 	f_\beta=\ \frac{1+70\chi_\beta}{1+80\chi_\beta} 
	\label{eq:29} \\
	\beta^\ast=\ \beta_i^\ast\left[1+\ \zeta^\ast F\left(M_t\right)\right], \qquad  \qquad  
	\beta=\ \beta_i\left[1-\frac{\beta_i^\ast}{\ \beta_i}\zeta^\ast F(M_t)\right]  \label{eq:30} \\
	\text{where}\quad 
	\beta_i^\ast=\beta_\infty^\ast\left[\frac{({4}/{15})+\ \left(Re_{\beta}\right)^4}{1+\ \left(Re_{\beta}\right)^4}\right] , 	\quad
	Re_{\beta}=\frac{Re_t}{R_\beta}, \quad R_\beta=8, \quad  \zeta^\ast=1.5, \quad \beta_\infty^\ast=0.09
	\nonumber\\
%\end{gather}
%
%\begin{gather}
%	f_\beta=\ \frac{1+70\chi_\beta}{1+80\chi_\beta}, \qquad
%	\beta=\ \beta_i\left[1-\frac{\beta_i^\ast}{\ \beta_i}\zeta^\ast F(M_t)\right]  
%	\label{eq:31}  \\
%	\\
%	\text{where}\quad 
	\chi_k\ \equiv\ \frac{1}{\omega^3}\frac{\partial k\ }{\partial x_j}\frac{\partial\omega\ }{\partial x_j}, \quad
	\chi_\beta=\ \left|\frac{\Omega_{ij}\Omega_{jk}S_{ki}}{\left(\beta_\infty^\ast\omega\right)^3}\right| \nonumber \\
	\Omega_{ij}=\ \frac{1}{2}\ \left[\frac{\partial u_i}{\partial x_j}-\frac{\partial u_j}{\partial x_i}\ \right], \quad
	S_{ki}=\ \frac{1}{2}\ \left[\frac{\partial u_i}{\partial x_k}+\frac{\partial u_k}{\partial x_i}\ \right]
	\label{eq:32}  \nonumber
	%	\\
	%	S_{ki}=\ \frac{1}{2}\ \left[\frac{\partial u_i}{\partial x_k}+\frac{\partial u_k}{\partial x_i}\ \right], \quad
	%	\beta=\ \beta_i[1-\frac{\beta_i^\ast}{\ \beta_i}\zeta^\ast F(M_t)]
	%	\label{eq:32a}  
\end{gather}
%\\
\remove{In the dissipation term  $\rho\beta f_\beta\omega^2$ of the transport equation of $\omega$ in \mbox{\eqn(\ref{eq:24})}}
\removeEq{
\begin{gather*}
	f_\beta=\ \frac{1+70\chi_\beta}{1+80\chi_\beta}, \qquad
	\beta=\ \beta_i\left[1-\frac{\beta_i^\ast}{\ \beta_i}\zeta^\ast F(M_t)\right]  
	\label{eq:31}  \\
	\text{where}\quad 
	\chi_k\ \equiv\ \frac{1}{\omega^3}\frac{\partial k\ }{\partial x_j}\frac{\partial\omega\ }{\partial x_j}, 	\quad
	Re_{t\beta}=\frac{Re_t}{Re_\beta}, \quad Re_\beta=8, \quad  \zeta^\ast=1.5, \quad \beta_\infty^\ast=0.09 \nonumber \\
	\chi_\beta=\ \left|\frac{\Omega_{ij}\Omega_{jk}S_{ki}}{\left(\beta_\infty^\ast\omega\right)^3}\right|,\quad
	\Omega_{ij}=\ \frac{1}{2}\ \left[\frac{\partial u_i}{\partial x_j}-\frac{\partial u_j}{\partial x_i}\ \right], \quad
	S_{ki}=\ \frac{1}{2}\ \left[\frac{\partial u_i}{\partial x_k}+\frac{\partial u_k}{\partial x_i}\ \right]
	\label{eq:32}  \nonumber
	\\
	S_{ki}=\ \frac{1}{2}\ \left[\frac{\partial u_i}{\partial x_k}+\frac{\partial u_k}{\partial x_i}\ \right], \quad
	\beta=\ \beta_i[1-\frac{\beta_i^\ast}{\ \beta_i}\zeta^\ast F(M_t)]
	\label{eq:32a}  
\end{gather*}}
%
%\subsubsection*
\textit{(d) Compressibility correction: }
\replace{$F\left(M_t\right)$ is a}{The} compressibility function \add{$F(M_t)$ appearing in \eqn(\ref{eq:30}) is given by}
\begin{gather}
	F\left(M_t\right) = 
	\begin{cases}
		0 &M_t<M_{t0}\\
		M_t^2-M_{t0}^2 &M_t>M_{t0}
	\end{cases}
	\label{eq:33} \\
	\text{where}\qquad 
	M_t^2\ \equiv\ \frac{2k}{a^2}, 
	\quad
	M_{t0}=0.25, \qquad 
	a = \sqrt{\gamma RT}
	\label{eq:34} 
\end{gather}
where $\gamma$ is the adiabatic index, $R$ is the gas constant, $a$ is the speed of sound, and $T$ is the temperature. 
\\
In the high Reynolds number form of the $k-\omega$ model, $\beta_i^\ast=\ \beta_\infty^\ast$\add{, and $\alpha^\ast=\ \alpha_\infty^\ast=1$}. In the incompressible form, $\beta^\ast=\ \beta_i^\ast$.
\remove{The model constants are as follows:}
\removeEq{\begin{gather*}
	\alpha_\infty^\ast=1,\quad \alpha_0= \frac{1}{9},\quad \alpha^\ast=0.52,\quad \beta_\infty^\ast=0.09,\quad \beta_i=0.072,\quad R_\beta=8,\\ 
	R_k=6,\quad R_w=2.95,\quad \zeta^\ast=1.5,\quad M_{t0}=0.25,\quad \sigma_k=2.0,\quad \sigma_\omega=2.0
	\label{eq:35} 
\end{gather*}}
%
%--------------------------------------
\subsubsection{Shear stress transport (SST) $k-\omega$ model}
%--------------------------------------
%
The SST \remove{shear stress transport} $k-\omega$ model  \citep{Menter1994} effectively blends \remove{the formulation of} $k-\omega$ model in the near-wall region and $k-\epsilon$ model in the outer region of the flow. The turbulent kinetic energy ($k$) and specific dissipation rate ($\omega$) are obtained from the following equation: 
\begin{gather}
	\frac{\partial\rho k}{\partial t}+\ \frac{\partial}{\partial x_j}\left(\rho k{\overbar{u}}_j\right)=\ \frac{\partial}{\partial x_j}\left[\left(\mu+\ \frac{\mu_t}{\sigma_k}\right)\frac{\partial k}{\partial x_j}\right]+2\mu_t{\overbar{S}}_{ij}\cdot{\overbar{S}}_{ij}-\rho\beta^*f_{\beta^*} k\omega	
	\label{eq:36} \\
	\frac{\partial\rho \omega}{\partial t}+\ \frac{\partial}{\partial x_j}\left(\rho \omega{\overbar{u}}_j\right)=\ \frac{\partial}{\partial x_j}\left[\left(\mu+\ \frac{\mu_t}{\sigma_\omega}\right)\frac{\partial \omega}{\partial x_j}\right]+\left(\frac{\alpha}{\nu_t}\right)2\mu_t{\overbar{S}}_{ij}\cdot{\overbar{S}}_{ij}-\rho\beta_i f_\beta k \omega^2	+ D_\omega
	\label{eq:37} 
\end{gather}
\remove{The isotropic eddy viscosity is computed from the following equation:}
\begin{gather}
	\mbox{where,}\qquad 
	\mu_t=\frac{\ \rho k}{\omega}\ {\left(\max\left[\frac{1}{\alpha^\ast},\ \frac{S_{ij}F_2}{\alpha_1\omega}\right]\right)^{-1}},
		\label{eq:38}\\
	\sigma_k= {\left(\frac{F_1}{\sigma_{k,1}}+\frac{1-F_1}{\sigma_{k,2}}\right)^{-1}},
	\qquad 
	\sigma_\omega= {\left(\frac{F_1}{\sigma_{\omega,1}}+\frac{1-F_1}{\sigma_{\omega,2}}\right)^{-1}}
	\label{eq:38a}\\
	F_1=\tanh{\left(\phi_1^4\right)},\qquad F_2=\tanh{\left(\phi_2^2\right)} 
	\label{eq:39} \\
	\phi_1=\min\left[\max(g_1, g_2), \frac{4\rho k}{\sigma_{\omega,2}D_{\omega}^+ y^2}\right], \qquad \phi_2=\max(2g_1, g_2)
	\label{eq:40} \nonumber\\
	g_1 = \frac{\sqrt k}{0.09\omega y},\quad g_2 = \frac{500\mu}{\rho y^2\omega},\qquad D_\omega^+=\max\left[2\rho\frac{1}{\rho_{\omega,2}\ }\ \frac{1}{\omega}\ \frac{\partial k}{\partial x_j}\ \frac{\partial\omega}{\partial x_j},\ {10}^{-10}\right]
	\label{eq:41}  \nonumber
	\end{gather}
Where $\alpha^\ast$ is defined in \eqn(\ref{eq:26}), $F_1$ and $F_2$ are blending functions, $y$ is the distance to the next surface, and $D_\omega^+$ is the positive portion of the cross-diffusion term ($D_\omega$). 
%
%\subsubsection*
\\
\textit{(a) Modelling turbulence production: }
%
%\subsubsection*{Production of $k$}
%
\replace{The turbulence production term $2\mu_t{\overbar{S}}_{ij}\cdot{\overbar{S}}_{ij}$ in \mbox{\eqn(\ref{eq:36})} represents the production of $k$ is the same as the standard $k-\omega$ model.%\\
%
%\subsubsection*{Production of $\omega$}
%
The term $({\alpha}/{\nu_t})2\mu_t{\overbar{S}}_{ij}\cdot{\overbar{S}}_{ij}$ found \mbox{(in \eqn\ref{eq:37}}) represents the production of $\omega$,} {The terms $2\mu_t{\overbar{S}}_{ij}\cdot{\overbar{S}}_{ij}$ {(in \eqn \ref{eq:36}}) and $({\alpha}/{\nu_t})2\mu_t{\overbar{S}}_{ij}\cdot{\overbar{S}}_{ij}$ in \eqn(\ref{eq:37}) represent the production of $k$ and $\omega$, respectively, similarly as in  the standard $k-\omega$ model. The coefficient} $\alpha$ is given by \eqn(\ref{eq:28}) \add{with} $\alpha_\infty$ \replace{is given by}{defined as}
\begin{gather}
	\alpha_\infty=[F_1\alpha_{\infty,1}+\left(1-F_1\right)\alpha_{\infty,2\ }] 
		\label{eq:42} \\ 
	\quad\text{where,}\quad 
	\alpha_{\infty,j}=\left(\frac{\beta_{i,j}\ }{\beta_\infty^\ast} -  \frac{\kappa^2}{\sigma_{\omega,j}\sqrt{\beta_\infty^\ast}}\right),
	\quad \kappa=0.41 \nonumber
\end{gather}
%
%\subsubsection*
\textit{(b) Modelling turbulence  dissipation: }
%
%\subsubsection*{Dissipation of $k$}
%
\replace{The dissipation of $k$ in the SST $k-\omega$ model is defined similarly as in the standard $k-\omega$ model using the term $\rho\beta^\ast k\omega$ in \mbox{\eqn(\ref{eq:36})}. In the SST K-omega model $f_{\beta^\ast}$ is constant and is equal to 1. 
%
%\subsubsection*{Dissipation of ${\omega}$}
%
The dissipation of $\omega$ in the SST $k-\omega$ model is also defined in a similar way as in the standard $k-\omega$ model using the term $\rho\beta_i\omega^2$ in \mbox{\eqn(\ref{eq:37})}.  The $f_\beta$ is constant and is equal to 1.}
{In the SST $k-\omega$ model,  the dissipation of $k$ and $\omega$ is defined similarly as in the standard $k-\omega$ model using the term $\rho\beta^\ast f_{\beta^\ast}k\omega$ in \eqn(\ref{eq:36}) and $\rho\beta_i f_\beta k\omega^2$ in \eqn(\ref{eq:37}), respectively, with $f_\beta = f_{\beta^\ast} = 1$.}
The $\beta_i$ is given by the following equation
\begin{gather}
\beta_i=F_1\beta_{i,1}+\left(1-F_1\right)\beta_{i,2}	
\label{eq:43} 
\end{gather}
\add{where,} $F_1$ is obtained from \eqn(\ref{eq:39}). 
%
%\subsubsection*
\\
\textit{(c) Cross diffusion modification: }
\replace{The SST $k-\omega$ model is based on both the standard $k-\omega$ model and the standard $k-\epsilon$ model. To blend these two models, the standard $k-\epsilon$ model is transformed into equations based on the $k-\omega$ model which leads to the introduction of a cross-diffusion term defined by the following term in the  \mbox{\eqn(\ref{eq:37})}.} {In the SST $k-\omega$ model, the cross-diffusion term ($D_\omega$ in \eqn \ref{eq:37}) results from the transformation of the standard $k-\epsilon$ model into the equations based on standard $k-\omega$ model to blend the standard $k-\omega$ and standard $k-\epsilon$ models. It is defined as follows.}
\begin{gather}
	D_\omega=2\left(1-F_1\right)\rho\frac{1}{\omega\sigma_{\omega,2}}\frac{\partial k}{\partial x_j}\frac{\partial\omega}{\partial x_j}
	\label{eq:44} 
\end{gather}
%
%\subsubsection*
\textit{(d) Model constants: }
The model constants are as follows: 
$\sigma_{k,1}=1.176,\ {\ \sigma}_{\omega,1}=\ 2.0,\ \sigma_{k,2}=1.0,\ \sigma_{\omega,2}=1.168,\ a_1=0.31,\ \beta_{i,1}=0.075,\ \beta_{i,2}=0.0828$.
All other  model constants (i.e., $\alpha_\infty^\ast,\ \alpha_\infty,\ \alpha_0,\beta_\infty^\ast,\ R_\beta,\ R_k,\ R_\omega,\ \zeta^\ast,  M_{t0}\ $) in SST $k-\omega$ model have the same values as the standard $k-\omega$ model. 
%
%--------------------------------------
\subsubsection{Linear Pressure Strain - Reynolds Stress Model (LPS-RSM)}
%--------------------------------------
%
Abandoning the isotropic eddy viscosity hypothesis, the RSM closes the RANS equations by solving the transport equation of Reynolds stresses ($R_{11}, R_{22}\ ,\ R_{33}\ ,\ R_{12},\ R_{13},\ R_{23}$) alongside the equation for the dissipation rate. Thus, \add{for the 3D flow,} seven additional equations \replace{are required to be solved for the 3D flow addition to}{with} the mean flow equations (\add{i.e., }continuity and momentum equations) \add{must be solved}. 
%
%\subsubsection*{(a) RSM Equations}
%
\\
\replace{Reynolds Stress equation models}{RSM Equations} are based on \replace{the utilization of Reynolds Stress Transport Equations. The}{utilizing the following} transport equation for \remove{the transport of} Reynolds stresses $R_{ij}=\langle \overbar{u_i^\prime\ u_j^\prime}\rangle$.
\begin{gather}
{\underbrace{\frac{\partial}{\partial t}\left(\rho R_{ij}\right)}_{ \text{local time derivative}}}
+ {\underbrace{\frac{\partial}{\partial x_k}\left(\rho u_kR_{ij}\right)}_{C_{ij}\ \equiv\  \text{Convection term}}} =
-{\underbrace{\frac{\partial}{\partial x_k}\left[\rho\overbar{u_i^\prime u_j^\prime u_k^\prime}+\overbar{p^\prime(\delta_{kj}u_i^\prime+\delta_{ik}u_j^\prime)}\ \right]}_{D_{T,ij}\ \equiv\  \text{Turbulent Diffusion}}} \nonumber\\
+ \underbrace{\frac{\partial}{\partial x_k}\left[\mu\frac{\partial}{\partial x_k}(\overbar{u_i^\prime u_j^\prime})\right]}_{{D_{M,ij}}\equiv\  \text{Molecular Diffusion}}
- {\underbrace{\rho\left(\overbar{u_i^\prime u_k^\prime}\frac{\partial u_j}{\partial x_k}+\ \ \overbar{u_j^\prime u_k^\prime}\frac{\partial u_i}{\partial x_k}\right)\ }_{P_{ij}\equiv\ \text{Stress Production}}}
\nonumber\\
+ \ {\underbrace{p\prime\left(\overbar{\frac{\partial{u^\prime}_i}{\partial x_j}+\ \frac{\partial{u^\prime}_j}{\partial x_i}}\right)}_{\phi_{ij}\equiv\ \text{Pressure Strain}}}
- {\underbrace{2\mu\overbar{\frac{\partial{u^\prime}_i}{\partial x_k}\ \frac{\partial{u^\prime}_j}{\partial x_k}}\ }_{\epsilon_{ij}\ \equiv\ \text{Dissipation}}}
	\label{eq:46} 
\end{gather}
In \eqn(\ref{eq:46}), \add{while} the terms $D_{M,ij}$ \add{and} $P_{ij}$ require no extra modelling\replace{. While}{,} the other three terms $D_{T,ij}$, $\phi_{ij}$ and $\epsilon_{ij}$ need to be modelled to close the equation. 
The turbulent diffusion term ($D_{T,ij}$) is modelled \replace{using the following equation}{as}
\begin{gather}
	D_{T,ij}=\ \frac{\partial}{\partial x_k}\left(\frac{\mu_t}{\sigma_k}\ \frac{\partial\overbar{u_i^\prime u_j^\prime}}{\partial x_k}\right)
	\label{eq:47} 
\end{gather}
where, $\sigma_{k\ \ }=\ 0.82$ \citep{Lien1994}. The turbulent viscosity ($\mu_t$) is computed using \eqn(\ref{eq:10}).  
%\\
The tensor ($\epsilon_{ij}$) is modelled using 
	\begin{gather}
		\epsilon_{ij}=\rho\epsilon\frac{2}{3}\ \delta_{ij}\left(1+2\frac{k}{a^2}\right)       
		\label{eq:48} 
	\end{gather}
where $a$ is the speed of sound (\eqn\ref{eq:34}),  
%\\
the scalar dissipation rate $\epsilon$ is computed using \remove{an equation similar to that used in the standard $k-\epsilon$ model} (\eqn\ref{eq:8}), and turbulent kinetic energy ($k$) is obtained by taking the trace of Reynolds stress tensor. 	
		\begin{gather}
		k=\ \frac{1}{2}\overbar{{u^\prime}_i{u^\prime}_i}    
		\label{eq:49} 
	\end{gather}
In this study, \remove{the Linear Pressure Strain Reynolds Stress Model (}LPS-RSM\replace{)} {\citep{Gibson1978,Fu1987,Launder1989}} is employed \replace{that models}{to model} the pressure strain term ($\phi_{ij}$)\remove{, its formulation is based on the proposal of \mbox{\citep{Gibson1978,Fu1987,Launder1989}}}.  
 The model \replace{uses the following decomposition to model}{decomposes the pressure strain} ($\phi_{ij}$) \add{into the slow pressure strain ($\phi_{ij,1}$), rapid pressure strain ($\phi_{ij,2}$), and wall reflection strain ($\phi_{ij,w}$)} as follows.
\begin{gather}
	\phi_{ij}= \phi_{ij,1}+ \phi_{ij,2}+ \phi_{ij,w}
	\label{eq:50} 
\end{gather}
\remove{where $\phi_{ij,1}$ is called the slow pressure strain term, $\phi_{ij,2}$ is known as rapid pressure strain term and $\phi_{ij,w}$ is called the wall reflection term. The slow pressure term ($\phi_{ij,1}$)  is modelled as}
\removeEq{\begin{gather}
	\phi_{ij,1}=\ -C_1\rho\frac{\epsilon}{k}\left[\overbar{{u^\prime}_i{u^\prime}_i}-\ \frac{2}{3}\delta_{ij}k\right]
	\label{eq:51} 
\end{gather}}
\remove{where 	$C_1 =1.8$.} The slow pressure strain term ($\phi_{ij,1}$) \replace{can also be}{is} represented using the Reynolds stress anisotropy tensor ($b_{ij}$) as  
\begin{gather}
	\phi_{ij,1} = {-2C}_1\rho\epsilon b_{ij}
	\qquad\text{where}\qquad C_1 =1.8,\qquad 
	b_{ij}=\ -\frac{1}{2k}\left({-\overbar{{u^\prime}_i{u^\prime}_i}+\ \frac{2}{3}\delta_{ij}k}\right)
	\label{eq:52} 
\end{gather}
The rapid pressure strain term ($\phi_{ij,2}$)  is modelled as
\begin{gather}
	\phi_{ij,2}=\ -C_2\left[\left(P_{ij}-C_{ij}\right)-\ \frac{2}{3}\delta_{ij}\left(\frac{1}{2}P_{kk}-\frac{1}{2}C_{kk}\right)\right] \qquad\text{where}\qquad C_2 =0.60\qquad 
	\label{eq:53} 
\end{gather}
where \remove{$C_2 = 0.60$,} $P_{ij}$ and $C_{ij}$ are defined in \eqn(\ref{eq:46}).  
\\
The wall reflection term ($\phi_{ij,w}$) redistributes normal stress near the wall. It tends to dampen the normal stress perpendicular to the wall while enhancing the stresses parallel to the wall. The term is modelled as 
\begin{gather}
	\phi_{ij,w}\ \equiv C_1^\prime\frac{\epsilon}{k}\ \left(\overbar{u_k^\prime u_m^\prime}n_kn_m\delta_{ij}-\ \frac{3}{2}\overbar{u_i^\prime u_k^\prime}n_jn_k-\ \frac{3}{2}\overbar{u_j^\prime u_k^\prime}n_in_k\right)\frac{C_lk^\frac{3}{2}}{\epsilon d} \nonumber\\
	+C_2^\prime\left(\phi_{km,2\ }n_kn_m\delta_{ij}-\frac{3}{2}\phi_{ik,2}n_jn_k-\ \frac{3}{2}\phi_{jk,2}n_in_k\right)\ \frac{C_lk^\frac{3}{2}}{\epsilon d}
	\label{eq:54} 
\end{gather}
where $C_1^\prime=0.5$,\ $C_2^\prime=0.3$, $C_l=({C_\mu^{\frac{3}{4}}}/{\kappa})$,  $C_\mu=0.09$, $n_k$ is the $x_k$ component of the unit normal to the wall, $d$ is the normal distance to the wall, and $\kappa = 0.4187$ is the von Karman constant. 
\begin{figure}[!b]
	\centering
	\subfigure[Front view]{\includegraphics[width=0.8\linewidth]{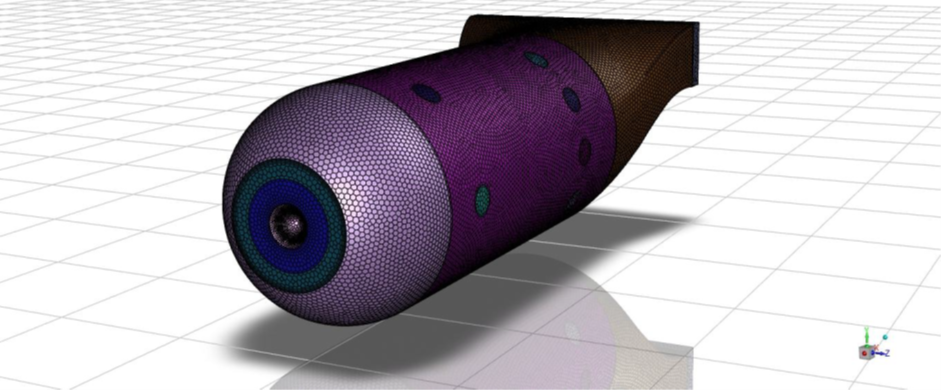}\label{fig:2}}
	\subfigure[A sectional view]{\includegraphics[width=0.8\linewidth]{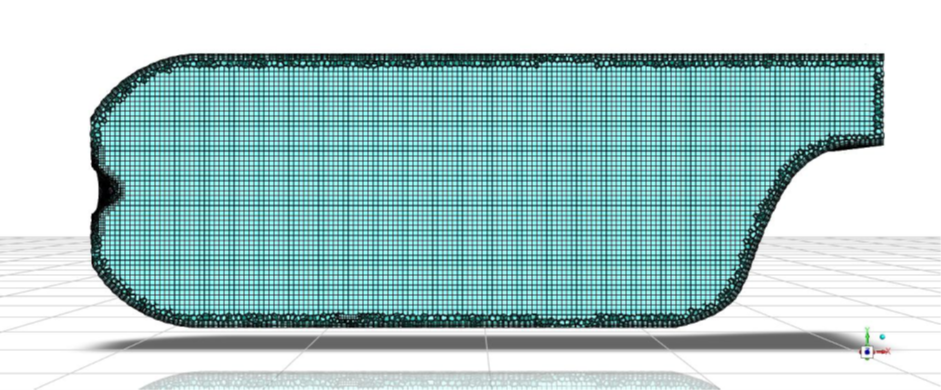}\label{fig:3}}
	\caption{Schematics of the mesh.\label{fig:3a}}
\end{figure}
%
%
%--------------------------------------
\section{Numerical Approach}
%--------------------------------------
%
In this study, the \replace{governing}{model} equations are solved using the ANSYS Fluent 2022 R1 (student version) solver, which employs the finite volume method  \citep{Versteeg2007}. Given the quasi-steady nature of the flow \remove{regime} \citep{Heitor1986}, \add{the} steady-state simulations have been \replace{conducted}{performed}. The pressure-velocity coupling \replace{was}{is} achieved \replace{through the utilization of}{by utilizing} a coupled algorithm \citep{Ghobadian2007}. \replace{The discretization of mass and momentum employed the second-order upwind scheme, while the PRESTO scheme \mbox{\citep{ Patankar1980}}, which is specifically recommended for swirling flow \mbox{\citep{Cellek2018}}, was used to discretize the pressure.}{PRESTO scheme \citep{ Patankar1980}, recommended explicitly for swirling flow \citep{Cellek2018}, is used to discretize the pressure, and all other terms are discretized using the second-order upwind scheme. The convergence criteria of $10^{-4}$ is taken for all the parameters.} 
\\
\replace{The mesh generation is also performed using ANSYS Fluent 2022 R1 Student version. The mesh type is poly-hexacore, a hybrid type of}{ANSYS has been used to generate a hybrid poly-hexacore} mesh containing a hexahedral core and polyhedral outer shell (see \fig\ref{fig:3a}). Three layers of polyhedral prism-type cells on the walls have been generated to capture the boundary layer well (see \fig\ref{fig:3}). The swirler effective area is computed using the approach proposed by \citet{McGuirk1992}\add{,} and the boundary conditions (\tab\ref{tab:3}) are modeled using the approach proposed by \citet{Crocker1997}, similar to previous studies \remove{conducted by }\citep{Jones2002,diMare2004,Oefelein2006,Veynante2009}. 
\\
 \replace{In t}{T}he mesh convergence study\replace{,}{is performed to assess} the influence of grid density on velocity and turbulence fields \remove{is assessed. Mesh Convergence simulations are performed} using the standard $k-\epsilon$ model \citep{Launder1972}. \replace{The simulation of n}{N}ear-wall turbulence {was conducted}{is simulated} using the enhanced wall treatment \remove{Method} (EWT)\remove{. EWT is a} modeling approach that combines linear (i.e., laminar) and logarithmic (i.e., turbulent) laws of the wall using a \replace{joining }{blending} function \citep{Kader1981}.  
\replace{Following the guidelines \mbox{\citep{Celik2008}}, the}{The mesh refinement} ratio $r =h_{\text{coarse}}/h_{\text{fine}}$ \replace{of Max}{for maximum} cell length is maintained above 1.3 \citep{Celik2008}.  The \add{characteristics of the} meshes used \add{in this study} are described in \tab\ref{tab:2}. \remove{The boundary conditions used are described in \mbox{\tab\ref{tab:3}}.}
\begin{table}[!htbp]
	\centering
	\caption{Mesh characteristics.}\label{tab:2}
	\scalebox{1}
	{
		\begin{tabular}{|c|c|c|c|}
			\hline
			Mesh & Elements & Maximum cell length, $h$ (m)	& $r =h_{\text{coarse}}/h_{\text{fine}}$\\\hline
			1	& 113599 & 0.003	& -- \\\hline
			2	& 210417 & 0.00225	& 1.333\\\hline
			3	& 429720 & 0.0016	& 1.40625\\\hline
		\end{tabular}
	}
\end{table}
%
%--------------------------------------
\section{Results and Discussion} % Numerical Analysis 
%--------------------------------------
%
\add{Before presenting the assessment of the turbulence model in predicting the isothermal flow in a realistic can-type gas turbine combustor geometry (refer \fig\ref{fig:1}), the mesh independence study has first been performed and discussed. Subsequently, the reliability and accuracy of the present modeling approach are presented by comparing and validating the present predictions with the limited experimental results available in the literature \citep{Heitor1986,Heitor1985}. The velocity field is normalized using the bulk velocity ($U_b=31$ m/s) \citep{Heitor1986,Heitor1985} and radial position is normalized by the internal radius ($R_c = 37.5$ mm)  as $r^\ast = r/R_c$.} 
\subsection{Mesh independence study}
\begin{figure}[!tb]
	\centering
	\subfigure[Axial Velocity]{\includegraphics[width=0.48\linewidth]{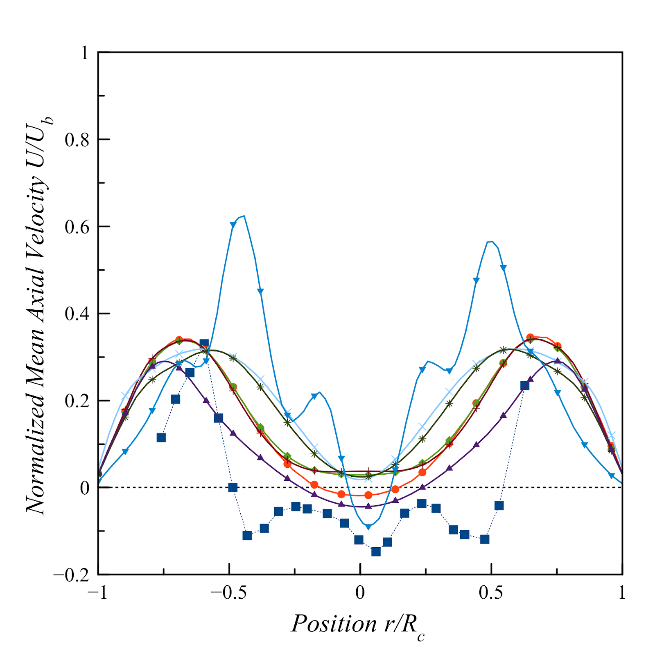}\label{fig:4a}}
	\subfigure[Transverse Velocity]{\includegraphics[width=0.48\linewidth]{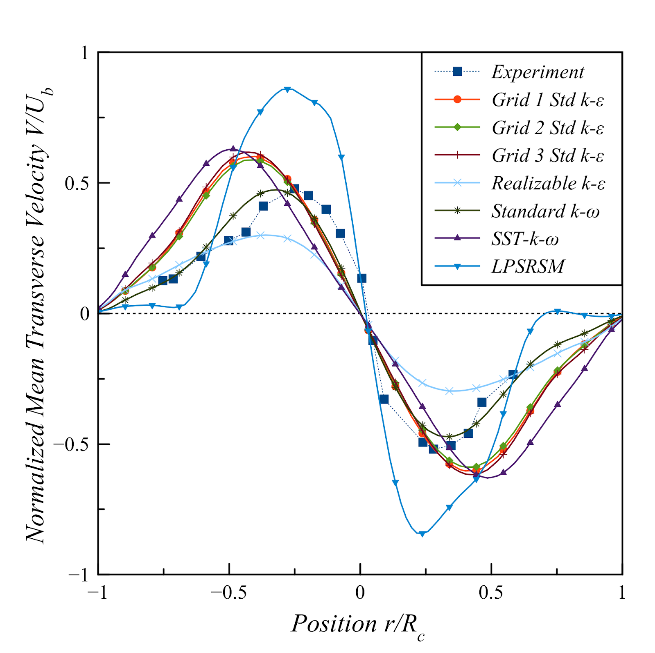}\label{fig:4b}}
	\subfigure[Turbulent Kinetic Energy]{\includegraphics[width=0.48\linewidth]{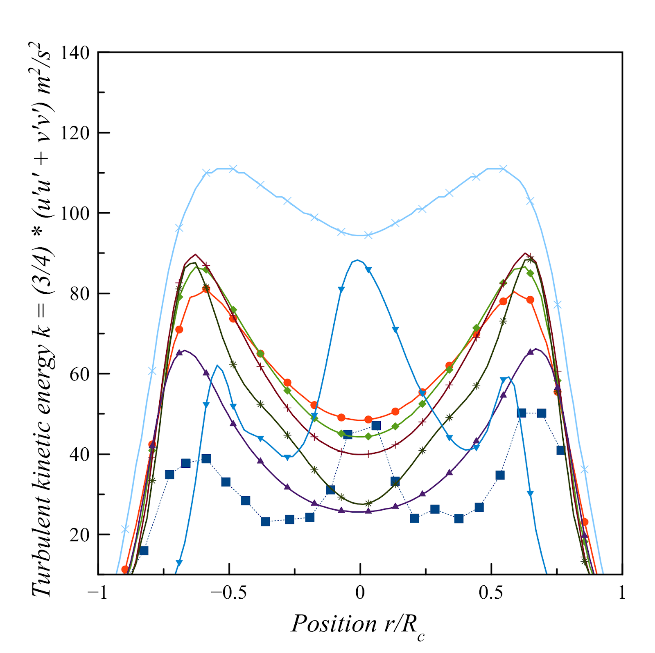}\label{fig:4c}}
	\subfigure[Shear Stress]{\includegraphics[width=0.48\linewidth]{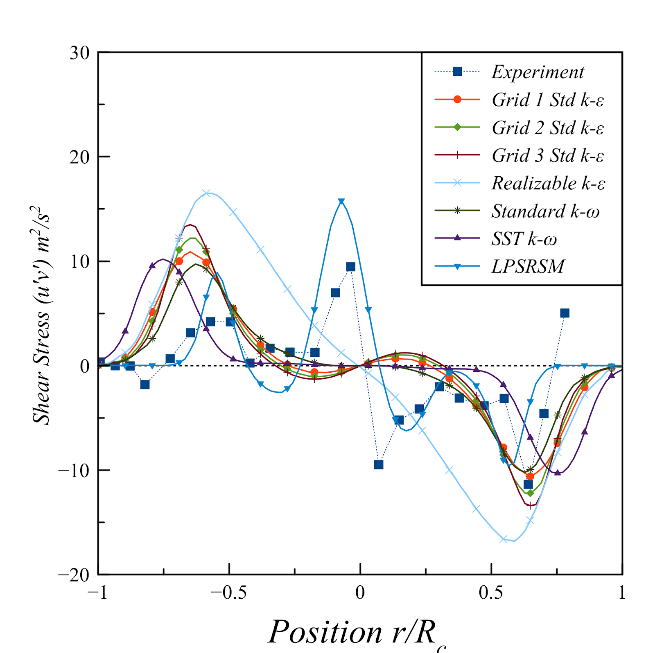}\label{fig:4d}}
	\caption{Velocity and turbulence characteristics at axial position $z = 50$ mm on primary holes plane at isothermal flow conditions.\label{fig:4}}
\end{figure}

\begin{figure}[!tb]
	\centering
	\subfigure[Axial Velocity]{\includegraphics[width=0.48\linewidth]{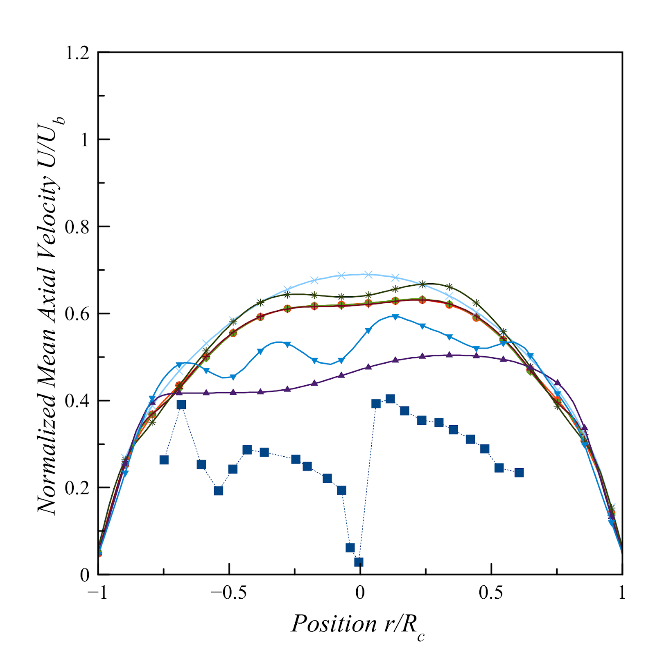}\label{fig:5a}}
	\subfigure[Transverse Velocity]{\includegraphics[width=0.48\linewidth]{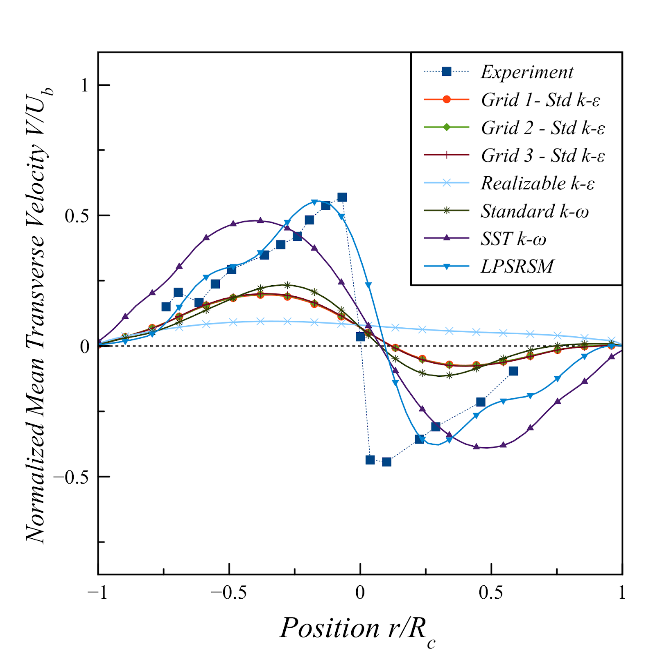}\label{fig:5b}}
	\subfigure[Turbulent Kinetic Energy]{\includegraphics[width=0.48\linewidth]{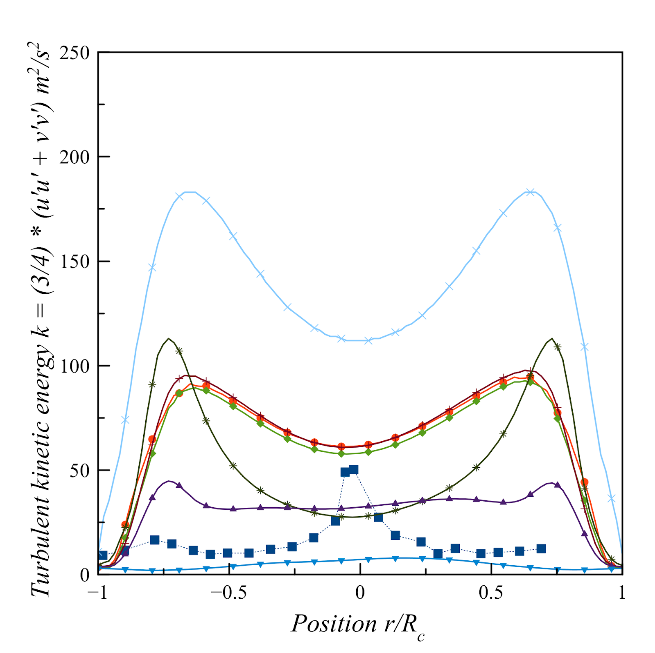}\label{fig:5c}}
	\subfigure[Shear Stress]{\includegraphics[width=0.48\linewidth]{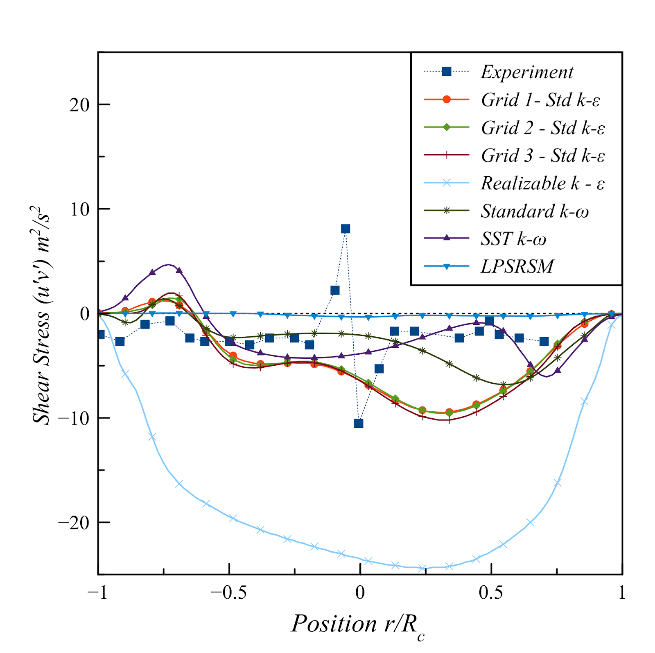}\label{fig:5d}}
	\caption{Velocity and turbulence characteristics at axial position $z = 130$ on dilution  holes plane at isothermal flow conditions.\label{fig:5}}
\end{figure}

The predicted velocity \add{and turbulence} characteristics \add{obtained using the three different grids  (\tab\ref{tab:3}) and standard $k-\epsilon$ model} are compared in \figs\ref{fig:4} and \ref{fig:5}.
%	
%\replace{T}{First, t}he predicted velocity \add{and turbulence} characteristics are compared \add{ and validated} \remove{with that of the experimental in} \figs\ref{fig:4} and \ref{fig:5} \add{ with the existing experimental results \citep{Heitor1986,Heitor1985}. Both \figs\ref{fig:4} and \ref{fig:5} include the predictions using various turbulence models (i.e., standard $k-\epsilon$, realizable $k-\epsilon$, standard $k-\omega$, SST $k-\omega$, and LPS-RSM), and different meshes (\tab\ref{tab:3}) with standard $k-\epsilon$ model.}  
\remove{The predicted velocity profiles are normalized using the bulk velocity which is $U_b=31$ m/s \mbox{\citep{Heitor1986,Heitor1985}}.} 
%\\
\replace{The}{\fig\ref{fig:4a} shows qualitatively similar} predicted \add{axial} velocity  \add{($U/U_b$)}  profiles at the position $z = 50$ mm (on the primary holes plane) \replace{in  {\fig\ref{fig:4a}} on grids 1, 2, and 3 show similarity}{for all three grids}. \replace{However, while observing the predicted mean axial velocity profile prediction of grid 1, it can be seen that it is reaching a negative value at the radial position $r/R_c\  = -0.25$ to 0.25.}{The mean axial velocity profile for grid 1, however, displays the negative values for the radial position $-0.25\le r^\ast \le 0.25$.} \replace{This is mainly attributed to the lower density of mesh elements in grid 1 which has resulted}{This attribute is due to the lower mesh density in grid 1, resulting } in numerical diffusion errors \citep{Versteeg2007}. 
%\\
Furthermore, \replace{in {\fig\ref{fig:4b}}, the transverse velocity  predictions at the position $z = 50$ mm indicate minimal variations for grids 1, 2, and 3.}{the grid size has indicated a minimal variation in the mean transverse velocity ($V/U_b$) predictions at the position $z = 50$ mm (on the primary holes plane) in \fig\ref{fig:4b}.}
\\
On observing the \replace{predicted T}{t}urbulent kinetic energy \add{($k$)} for the same position ($z = 50$ mm) \add{on primary holes plane}, predicted results in \fig\ref{fig:4c} show a steeper profile for \replace{obtained using a }{the} dense grid than the course grid. However, \replace{it can be stated that variations between grids 1, 2 and 3 are not immense.}{the influence of grid size on the turbulent kinetic energy is insignificant.}  This is because \remove{as the mesh element size becomes smaller,} the magnitude of the mean velocity gradient increases \add{with refinement in the mesh (i.e., decreasing mesh element size)}, resulting in the observed variation\replace{. This behavior has also previously been observed in the author’s previous research}{consistent with the literature} \citep{Kumar2017,Kumar2021,Kumar2021f}. \replace{The same trend can be seen}{Similar trends are evident } for the shear stress in \fig\ref{fig:4d}, as the magnitude of the computed shear stress depends directly on the magnitude of mean velocity gradients in two-equation models (see \eqn\ref{eq:3}).
\\
\add{Subsequently, the influence of mesh size on the predicted velocity and turbulence characteristics at the position $z = 130$ mm (on the dilution holes plane) are analyzed and shown in \fig\ref{fig:5}.}  
\replace{While observing the predicted axial velocity at the position $z = 130$ mm (on the dilution holes plane) in {\fig\ref{fig:5a}}, it is apparent that predictions using grids 1, 2 and 3 are similar.}{Evidently, the mesh size has shown negligible influence on the axial velocity ($U/U_b$) and mean transverse velocity  ($V/U_b$) in \figs\ref{fig:5a} and \ref{fig:5b}.}
\replace{The presence of}{N}umerical diffusion is less pronounced at $z = 130$ mm \replace{compared to}{than} at $z = 50$ mm, where the swirling motion was highly dominant.
\remove{The mean transverse velocity profile on the same position, as shown in {\fig\ref{fig:5b}} also suggests 
similar profiles obtained for grids 1, 2 and 3.}
The mean transverse velocity profiles \remove{using the standard $k-\epsilon$ model} also indicate\add{s} diffusion of the swirl as the flow convects to the dilution holes using the \replace{used}{standard $k-\epsilon$} turbulence model. \replace{Further, t}{T}he predicted turbulent kinetic energy \add{($k$) the shear stress} profiles show a similar trend \add{at $z = 130$ mm} in \figs\ref{fig:5c} \add{and  \fig\ref{fig:5d}} as observed for the  position $z = 50$ mm \replace{and the same can be said for the predicted shear stress as shown in {\fig\ref{fig:5d}}}{ in \figs\ref{fig:4c} \add{and  \fig\ref{fig:4d}}}. 
\\
Therefore, it can be inferred that grid independence is being approached. To, ensure maximum 
accuracy, the authors have chosen to use `Grid 3' for further simulations and analysis.
%
%--------------------------------------
\subsection{Assessment of Turbulence Models} % Numerical Analysis 
%--------------------------------------
%
\add{In this section, the various turbulence models (i.e., standard $k-\epsilon$, realizable $k-\epsilon$, standard $k-\omega$, SST $k-\omega$, and LPS-RSM) used in this work are assessed by comparing the present computational results using Grid 3 with experimental studies \citep{Heitor1986,Heitor1985} for their suitability in predicting the isothermal flow in a realistic can-type gas turbine combustor geometry (refer \fig\ref{fig:1}).}
%
%\replace{T}{To present the reliability and accuracy of the modeling and simulation approaches used in this study, t}he predicted velocity \add{and turbulence} characteristics are compared \add{ and validated} \remove{with that of the experimental in} \figs\ref{fig:4} and \ref{fig:5} \add{ with the existing experimental results \citep{Heitor1986,Heitor1985}.} 
%{Both \figs\ref{fig:4} and \ref{fig:5} include the predictions using various turbulence models (i.e., standard $k-\epsilon$, realizable $k-\epsilon$, standard $k-\omega$, SST $k-\omega$, and LPS-RSM).}  
%\\
\subsubsection{Standard $k-\epsilon$ model predictions}\label{sec:skep}
\begin{figure}[!tb]
	\centering
	\subfigure[at $z = 50$ mm on primary holes plane]{\includegraphics[width=0.48\linewidth]{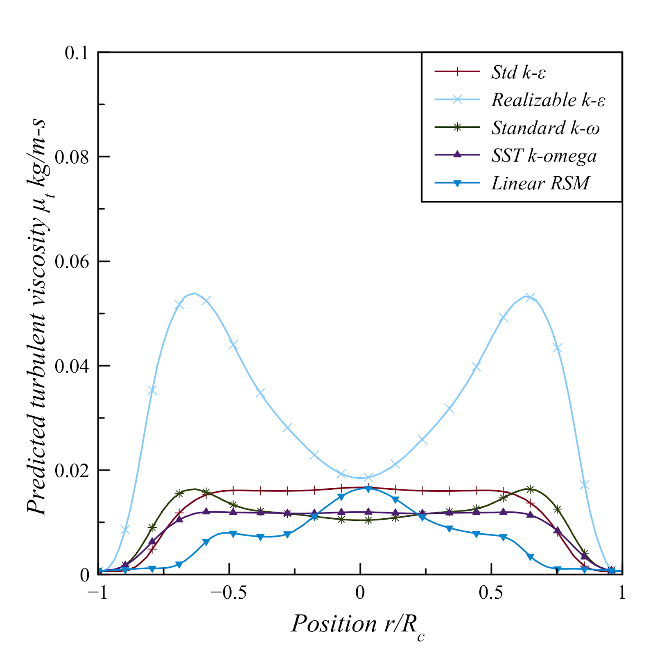}\label{fig:6}}
	\subfigure[at $z = 130$ mm on dilution holes plane]{\includegraphics[width=0.48\linewidth]{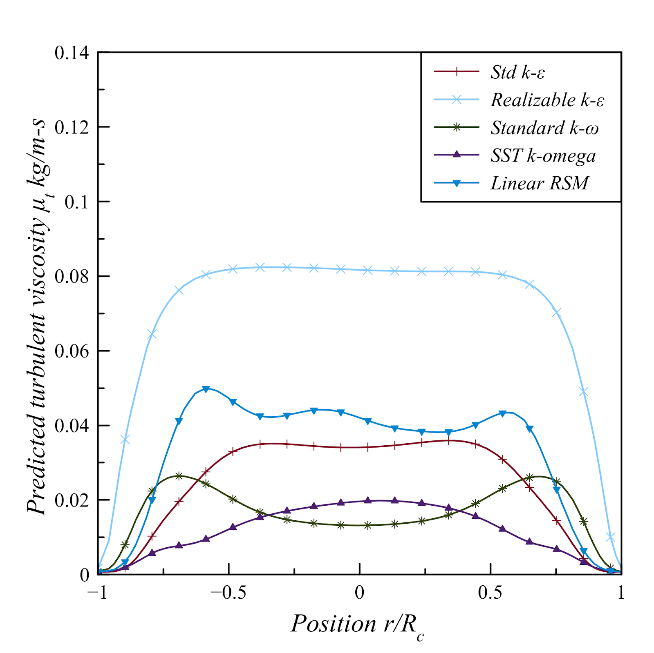}\label{fig:7}}
	\caption{Predicted turbulent viscosity at the axial position $z$ on primary and dilution holes plane at isothermal flow conditions.}
\end{figure}

\replace{On comparing the predicted axial velocity profile using the standard $k-\epsilon$ model using grid 3 on the location $z = 50$ mm on the dilution holes plane in {\fig\ref{fig:4a}}, it can be seen that the predicted axial velocity is not in the trend with the experimental measurement. The only position where it matches with the experimental values is at the radial position $r/R_c\ =\ -0.6$. The predicted axial velocity showing the non-negative axial velocity values indicates a lack of formulation present in the standard $k-\epsilon$ model to capture the central backflow which is represented in the experimental values in the radial position $r/R_c = -0.5$ to 0.5.}{The predicted axial velocity ($U/U_b$) obtained using the standard $k-\epsilon$ model has been compared with experimental measurements in \fig\ref{fig:4a} on the location $z = 50$ mm on the dilution holes plane. Broadly, the two results indicate an inconsistent trend, except at the location $r^\ast  = -0.6$ where the two values match. The non-negative predicted axial velocity suggests a lack of formulation in the standard $k-\epsilon$ model to capture the central backflow\add{,} which is evident in the experimental values in the radial position $-0.5\le r^\ast \le 0.5$. }
\replace{The standard $k-\epsilon$ model is based on the assumption of local isotropy which means that the turbulence is the same in all the directions at a given point. This assumption fails to accurately capture the complex flow patterns and vortical structures associated with swirling flows. This also leads to inadequate representation of the turbulent transport processes, including redistribution of momentum and the formation of secondary flows. The consequence of it can be seen in {\fig\ref{fig:4b}} where transverse velocity is underpredicted which indicates a lack of swirl in the vortex using the standard $k-\epsilon$ model.}{The standard $k-\epsilon$ model assumes that the local isotropy (i.e., turbulence is the same in all the directions at a given point) \citep{Tang2022}. This assumption leads to an inadequate representation of the turbulent transport processes, including the redistribution of momentum and the formation of secondary flows, and, thus,  inaccurately captures the complex flow patterns and vortical structures associated with swirling flows. Consequences are evident in \fig\ref{fig:4b} where transverse velocity  ($V/U_b$) is underpredicted, compared with experimental measurements, which indicates a lack of swirl in the vortex using the standard $k-\epsilon$ model. }
\replace{The predicted transverse magnitude switches from positive to negative at the centre position $r/R_c = 0.0$ in {\fig\ref{fig:4b}} indicating the position of the vortex core.  Corresponding to this, the eddy viscosity curve of the standard $k-\epsilon$ model in {\fig\ref{fig:6}} shows flatness from positions $r/R_c = -0.75$ to $r/R_c = 0.75$ showing eddy viscosity not being sensitive to the flow characteristics of the present case.}{The predicted transverse velocity switches from positive to negative at the centre ($r^\ast  = 0$) in \fig\ref{fig:4b}, indicating the position of the vortex core. Corresponding to this, the eddy viscosity curve of the standard $k-\epsilon$ model in \fig\ref{fig:6} shows flatness for $-0.75\le r^\ast  \le 0.75$ showing eddy viscosity not being sensitive to the flow characteristics of the present case. }
\\
\replace{On comparing the predicted turbulence kinetic energy using the standard $k-\epsilon$ model in {\fig\ref{fig:4c}} it can be noticed that turbulence is inaccurately predicted. The turbulence is not represented well using the model, in confined swirling flows, turbulence is concentrated and intensified due to the swirling motion and flow recirculation within the core which is observed in experimental values.}{The turbulent kinetic energy ($k$) using the standard $k-\epsilon$ model compared to the experimental results in \fig\ref{fig:4c} also indicates an inaccurate turbulence prediction. The standard $k-\epsilon$ model does not accurately represent the turbulence in confined swirling flows; turbulence is concentrated and intensified due to the swirling motion and flow recirculation within the core, as observed in experimental values.}
\replace{Besides, higher values of turbulence kinetic energy are also experimentally recorded near the wall due to the swirl which is additionally fed by primary jets entering the combustor from primary holes. The oversimplistic isotropic assumption led to higher turbulence at the outer edge of the vortex core,  symmetrically, predicted at the positions $r/R_c = -0.75$ and $r/R_c = 0.75$ which is predicted due to the swirl originating from the swirler and dilution jets, and lower values predicted }{Besides, higher experimental values of turbulence kinetic energy are recorded near the wall due to the swirl, additionally fed by primary jets entering the combustor from primary holes. The oversimplistic isotropic assumption led to higher turbulence at the outer edge of the vortex core,  symmetrically predicted at $r^\ast  =\pm 0.75$ which is predicted due to the swirl originating from the swirler and dilution jets, and lower values are observed} at the core around $r^\ast  = \pm 0.25$ \remove{$r^\ast  = -0.25$ and $r^\ast  = 0.25$} in \fig\ref{fig:4c}.
\\
\replace{In the confined swirling flows, the shear stress is usually high due to the strong velocity gradients, especially at the vortex core which is observable in the experimental measurement around the position $r/R_c = 0.0$  in {\fig\ref{fig:4d}}. Comparison with shear stress prediction using standard k-epsilon again indicates inaccuracy. The maximum shear stress values are obtained at the edge of the vortex at the positions $r/R_c = -0.75$ and $r/R_c = 0.75$ using the standard $k-\epsilon$ model, however, negligible values of shear stress are obtained at the centre position $r/Rc = 0.0$, again indicating the inability of the model to capture the flow dynamics of confined swirling flows inaccurately. The velocity vectors in {\fig\ref{fig:8a}} and streamlines of standard $k-\epsilon$ in {\fig\ref{fig:9a}} predictions also do not indicate flow reversal at the vortex core. The q-criterion shown in {\fig\ref{fig:10a}} also indicates that the vortex core progressively dilutes as the flow convects downstream towards dilution holes as mentioned above.}{In the confined swirling flows, the shear stress is usually high due to the strong velocity gradients, especially at the vortex core, which is observable in the experimental measurement around the position $r^\ast = 0$  in \fig\ref{fig:4d}. A comparison of experimental and computational results again indicates inaccuracy.  The maximum shear stress values are obtained at the edge of the vortex at $r^\ast = \pm 0.75$ using the standard $k-\epsilon$ model; however, negligible values of shear stress obtained at the centre ($r^\ast = 0$) are again indicating the inability of the model to capture the flow dynamics of confined swirling flows accurately. The velocity vectors in \fig\ref{fig:8a} and streamlines of standard $k-\epsilon$ in \fig\ref{fig:9a} predictions also do not indicate flow reversal at the vortex core. The $Q-$criterion shown in \fig\ref{fig:10a} also suggests that the vortex core progressively dilutes as the flow convects downstream towards dilution holes.}
\\
\replace{On observing the predicted axial velocity from standard k-epsilon simulation on dilution holes, in {\fig\ref{fig:5a}}  the maximum velocity magnitude is predicted at central positions from $r/R_c = -0.25$ to $r/R_c = 0.25$. The predicted transverse velocity plot in {\fig\ref{fig:5b}} shows underprediction, the transverse velocity switches from positive to negative at the centre Position $r/R_c = 0.0$ however, the underprediction points to the prediction of less swirl and also indicates that the vortex core diffuses as the flow convects downstream.}{On observing the predicted axial velocity using standard $k-\epsilon$ at $z=130$ mm on dilution holes, in \fig\ref{fig:5a}, the maximum velocity magnitude is predicted at central positions from $-0.25 \le r^\ast \le 0.25$. Further, the transverse velocity field in \fig\ref{fig:5b} shows underprediction to experimental results; the transverse velocity switches from positive to negative at the center ($r^\ast = 0$). However, the underprediction points to less swirl prediction and indicates that the vortex core diffuses as the flow convects downstream. }
\replace{The turbulence kinetic energy is overpredicted as shown in {\fig\ref{fig:5c}}, which can be said to be due to the prediction of diffusion of the vortex core which enhances turbulence kinetic energy due to its mixing with the surrounding fluid. Still, the slightly higher prediction of turbulence kinetic energy close to the combustor wall could be attributed to additional turbulence produced due to the inlet of air inlets at dilution holes and its mixing with the surrounding fluid. While observing the experimental turbulence kinetic energy plot in the same figure, it can be seen that the maximum turbulence kinetic energy is recorded at the centre indicating the maximum fluctuations and thus steepest velocity gradients.}{The turbulence kinetic energy is overpredicted, as shown in \fig\ref{fig:5c}, because of the prediction of diffusion of the vortex core, which enhances turbulence kinetic energy due to its mixing with the surrounding fluid. Still, the slightly higher prediction of turbulence kinetic energy close to the combustor wall could arise from the additional turbulence produced due to the air inlets at dilution holes and their mixing with the surrounding fluid. The experimental turbulence kinetic energy in \fig\ref{fig:5c} shows that the maximum value is recorded at the center, indicating the maximum fluctuations and thus the steepest velocity gradients. }
\replace{The shear stress is also inaccurately predicted as shown in {\fig\ref{fig:5d}}, the experimental values show the maximum magnitude of shear stress at the centre indicating the maximum magnitude of velocity gradients at the vortex core while shear stress is less pronounced and overall negative in the positions $r/R_c = -1.0$ to $r/R_c = -0.25$ and from $r/R_c = 0.25$ to $r/R_c = 1.0$. }{The shear stress is also inaccurately predicted, as shown in \fig\ref{fig:5d}; the experimental values show the maximum magnitude of shear stress at the center, indicating the maximum magnitude of velocity gradients at the vortex core, while shear stress is less pronounced and overall negative for ($-1 \le r^\ast \le -0.25$) and ($0.25\le r^\ast \le 1$). }
While the simulation results show \remove{a} substantially larger magnitude of shear stress\replace{all over except at the centre}{, except at the center, } indicating intensified mixing is predicted between the vortex and the surrounding fluid and with additional air entering the combustor through dilution holes, due to which higher and inaccurate turbulence kinetic energy is predicted at the same position\add{,} as above mentioned. The near eddy viscosity curve predicted on the dilution hole in \fig\ref{fig:7} from positions \replace{$r/R_c = -0.5$ to $r/R_c = 0.5$}{$-0.5 \le r^\ast \le 0.5$} also indicates \replace{the model }{that the standard $k-\epsilon$} model assumes a uniform level of turbulence in this region regardless of the flow characteristics. This further substantiates a lack of sensitivity to the anisotropic nature of the swirling flows and the model’s inability to \add{accurately} capture the dissipation of turbulence energy and corresponding changes in the flow dynamics\remove{accurately}.
%
%
%\\
\subsubsection{Realizable $k-\epsilon$ model predictions}
\begin{figure}[!htb]
	\centering
	\subfigure[Standard $k-\epsilon$]{\includegraphics[width=0.48\linewidth]{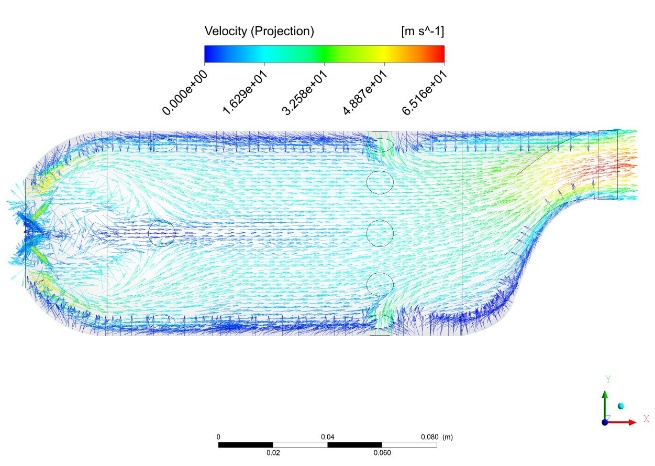}\label{fig:8a}}
	\subfigure[Realizable $k-\epsilon$]{\includegraphics[width=0.48\linewidth]{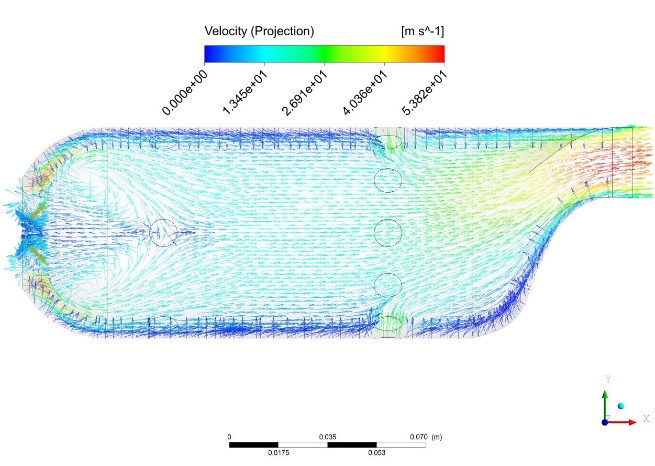}\label{fig:8b}}
	\subfigure[Standard $k-\omega$]{\includegraphics[width=0.48\linewidth]{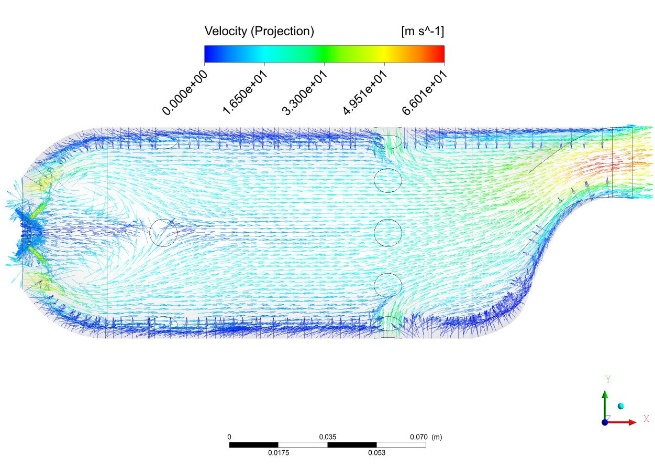}\label{fig:8c}}
	\subfigure[SST $k-\omega$]{\includegraphics[width=0.48\linewidth]{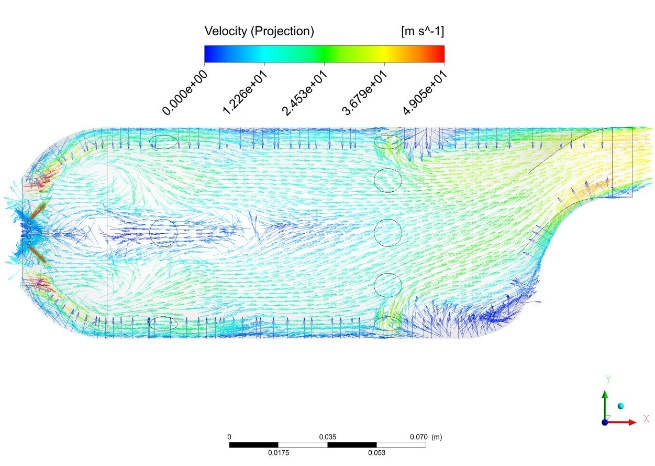}\label{fig:8d}}
	\subfigure[LPS-RSM]{\includegraphics[width=0.48\linewidth]{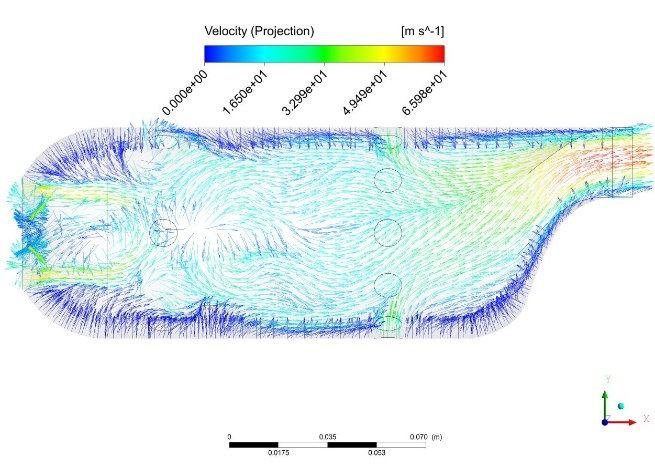}\label{fig:8e}}
	\caption{Comparison of velocity vectors predicted on X-Y plane using various tutbulence models (a) Standard $k-\epsilon$ model, (b) Realizable $k-\epsilon$ model, (c) Standard $k-\omega$, (d) SST $k-\omega$, (e) LPS-RSM.}\label{fig:8}
\end{figure}

\begin{figure}[!tb]
	\centering
	\subfigure[Standard $k-\epsilon$]{\includegraphics[width=0.48\linewidth]{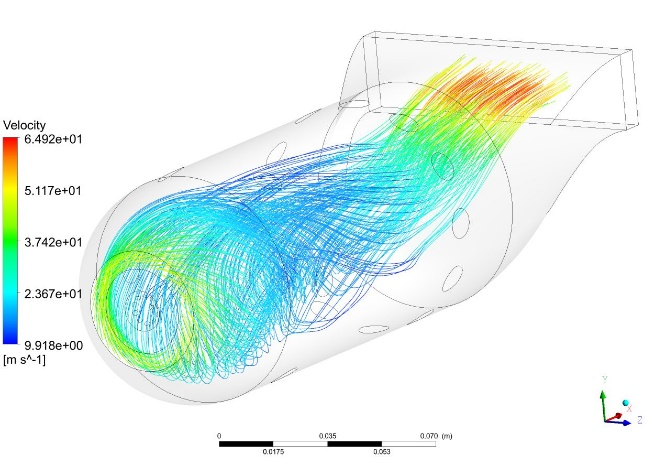}\label{fig:9a}}
	\subfigure[Realizable $k-\epsilon$]{\includegraphics[width=0.48\linewidth]{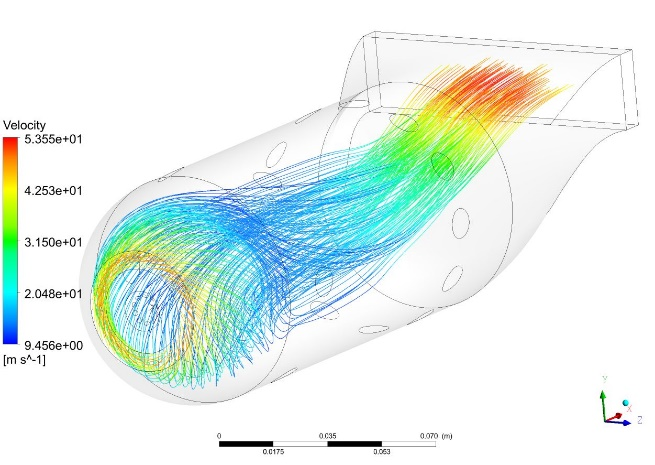}\label{fig:9b}}
	\subfigure[Standard $k-\omega$]{\includegraphics[width=0.48\linewidth]{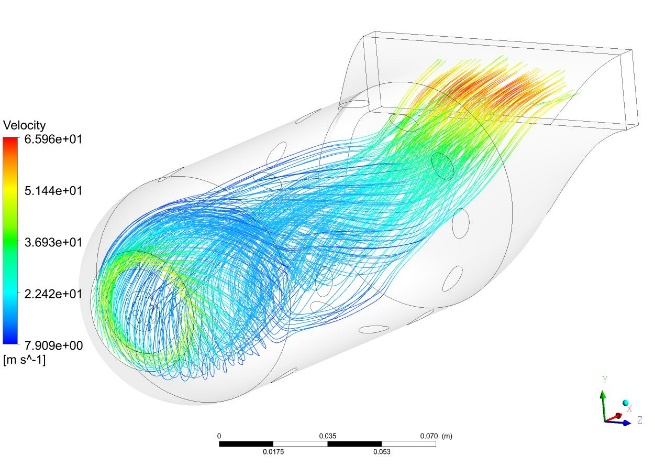}\label{fig:9c}}
	\subfigure[SST $k-\omega$]{\includegraphics[width=0.48\linewidth]{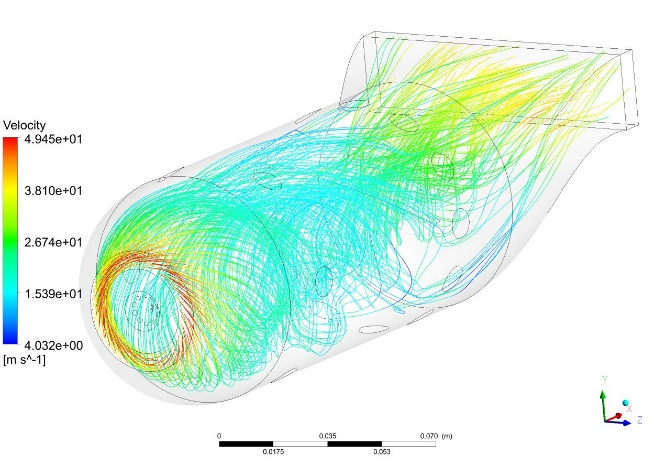}\label{fig:9d}}
	\subfigure[LPS-RSM]{\includegraphics[width=0.48\linewidth]{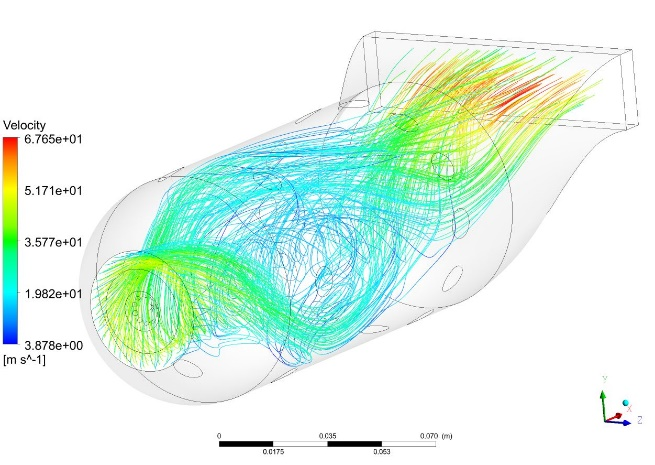}\label{fig:9e}}
	\caption{Comparison of velocity streamlines predicted using various tutbulence models (a) standard $k-\epsilon$, (b) realizable $k-\epsilon$, (c) standard $k-\omega$, (d) SST $k-\omega$, (e) LPS-RSM.}\label{fig:9}
\end{figure}

\replace{On analysing the results obtained using the realizable $k-\epsilon$ model on the primary hole plane at $z = 50$ mm, the axial velocity prediction is not in trend with experimental values and is overpredicted. }{\fig\ref{fig:4a} depicts that the axial velocity on the primary hole plane at $z = 50$ mm is overpredicted using the realizable $k-\epsilon$ model and inconsistent with the experimental values.}
\replace{The predicted axial velocity, obtained from the realizable $k-\epsilon$ model, exhibits non-negative values, hence, like standard $k-\epsilon$ the realizable $k-\epsilon$ model has also failed to capture the central backflow observed in the experimental data within the radial position range of $r/R_c = -0.5$ to 0.5. }{Similar to the standard $k-\epsilon$ model (refer section \ref{sec:skep}), the realizable $k-\epsilon$ model exhibits non-negative values for the axial velocity and failed to capture the central backflow observed in the experimental data within the radial position range of $0.5\le r^\ast  \le 0.5$.} 
\remove{This observed discrepancy again suggests that the formulation of the realizable $k-\epsilon$ model cannot accurately predict the complexity of confined swirling flows. }
%\\
The transverse velocity is also inaccurately predicted as shown in \fig\ref{fig:4b}. The predicted transverse velocity \replace{plot}{profile} reveals an underprediction of the transverse velocity which suggest\add{s} less swirl \remove{is} predicted in the vortex. 
\add{This observed discrepancy again suggests that the formulation of the realizable $k-\epsilon$ model cannot accurately predict the complexity of confined swirling flows.}  
\\
\replace{The realizable $k-\epsilon$ model shares the same transport equation for turbulence kinetic energy as the standard $k-\epsilon$ model while the transport equation of turbulent dissipation rate $\epsilon$ is derived using the transport equation of mean square of vorticity fluctuation {\citep{Tennekes2018}}.}{As noted earlier (refer section \ref{sec:rke}), the transport equation of turbulent dissipation rate ($\epsilon$) differ in the standard $k-\epsilon$ and realizable $k-\epsilon$ models, whereas that for turbulence kinetic energy ($k$) remains the same.}  
The model is \replace{called}{known as} realizable \replace{because}{as} it ensures the positivity of normal stress ($\overbar{u^\prime}^2>0$) \remove{which by definition is a positive quantity} and maintains the Schwarz inequality (refer \eqn\ref{eq:18}). 
\\
\replace{Schwarz inequality in the realizable $k-\epsilon$ model is used to bind the magnitude of shear stress in a fluid and to ensure that the stresses are realistic and physically meaningful.  The turbulence velocity formulation remains isotropic. The realizability is achieved by modifying the formulation of $C_\mu$ which is no longer constant like the standard $k-\epsilon$ model (refer {\eqn\ref{eq:19}}). This involves introducing additional terms and coefficients that depend on the strain rate and rotation rate (refer {\eqns\ref{eq:20}} and {\ref{eq:21}}) of the flows. These terms are aimed to assist in improving the turbulence-predicting ability of the model.  However, it can be stated that these additional terms and model constants are not calibrated for the confined swirling flows, apart from isotropic eddy viscosity assumption and due to these reasons simulation results obtained are even less accurate than the standard $k-\epsilon$ model. }{While the proportionality constant  ($C_\mu$) for the turbulent viscosity (refer \eqn\ref{eq:10})  has been modified by introducing additional terms/coefficients (\eqns\ref{eq:19} - \ref{eq:21}) to assist in improving the turbulence-predicting ability, apart from isotropic eddy viscosity assumption, it has not been calibrated for the confined swirling flows, and due to these reasons simulation results obtained using the realizable $k-\epsilon$ model are even less accurate than the standard $k-\epsilon$ model.} 
\\
The underprediction of transverse velocity indicates less swirl and a weak vortex core also {indicates}{demonstrates} that the model {is predicting}{predicts} a diffusion of vortex core even at the primary zone where experimental values indicate a very dominant swirl. 
The prediction of excessive diffusion of the vortex intensifies mixing with the surrounding fluid \replace{and}{, which} leads to the development of regions of high-velocity gradients and significant turbulence at the region of vortex interaction with the surrounding fluid or at the vortex's outer edge \citep{Davidson2015}. 
\replace{This leads to the prediction of high turbulent kinetic energy at this region as seen in {\fig\ref{fig:4c}} in positions $r/R_c = -0.75$ to $r/R_c = 0.75$. This has also led to the prediction of high-magnitude shear stress in this area shown in {\fig\ref{fig:4d}}.}{It produces high turbulent kinetic energy ($k$) and high-magnitude shear stress in the region $-0.75\le r^\ast \le 0.75$, as seen in \figs\ref{fig:4c} and \ref{fig:4d}, respectively.} 
\\
\replace{The realizable $k-\epsilon$ relates turbulent viscosity with turbulent kinetic energy and the turbulent dissipation rate through a closure relationship (refer {\eqn\ref{eq:10}}). The}{Further, the } turbulent viscosity \add{($\mu_t$, \eqn\ref{eq:10})} is directly proportional to the square of turbulent kinetic energy \add{($k$)} and inversely proportional to the turbulent dissipation rate \add{($\epsilon$)}.
However, the model has predicted lower turbulent kinetic energy and shear stress at the vortex core or positions \replace{$r/R_c = -0.25$ to $r/R_c = 0.25$}{$-0.25\le r^\ast\le 0.25$,} as seen in \figs\ref{fig:4c} and \ref{fig:4d}\replace{. This is}{,} in contrast to the experimental data\add{,} which indicate\add{s} high turbulent kinetic energy and shear stress at the vortex core.  
These discrepancies between the model\remove{’s} prediction and experimental data indicate that model\remove{’s} formulation \replace{is incapable of predicting}{cannot accurately predict} the intricacies of confined swirling flow\remove{ accurately}. 
Correspondingly, the velocity vectors \remove{in} (\fig\ref{fig:8b}) and streamlines \remove{in} (\fig\ref{fig:9b}), and $Q-$criterion \remove{in} (\fig\ref{fig:10b}) indicate a \replace{very small}{minimal} vortex core obtained using a realizable $k-\epsilon$ model.
\\
\replace{The predicted}{Furthermore, the} mean axial {and transverse} velocit\replace{y}{es} using the realizable $k-\epsilon$ model on the dilution hole plane \add{($z=130$ mm)} \replace{is not in trend}{are also overpredicted and inconsistent} with the experimental values \replace{and is overpredicted}{,} as seen in \figs\ref{fig:5a} \replace{.The predicted transverse velocity using the same model is also not in trend with the experimental of the same as seen in \mbox{\fig}}{and} \ref{fig:5b}. 
\replace{The predicted transverse velocity is positive throughout the plane, indicating no rotation and the vortex core originated upstream at the primary zone has completely diffused as the flow convected towards the dilution holes.}{The transverse velocity is positive throughout the plane, indicating no rotation, and the vortex core originated upstream at the primary zone ($z =50$ mm) has completely diffused as the flow convected towards the dilution holes ($z =130$ mm) plane.}
The predicted axial and transverse velocity profile\add{s} exhibit\remove{s} characteristics \remove{more} like a pipe flow\replace{ than the }{, different from }he expected flow profile of the combustor flow near the dilution holes \add{($z =130$ mm) plane}.
\\
\replace{On analysing the turbulent kinetic energy predictions by the realizable $k-\epsilon$ model in {\fig\ref{fig:5c}}, it can be seen that turbulent kinetic energy is clearly overpredicted and completely not in trend with the experimental data.}{The realizable $k-\epsilon$ model also overpredicts the turbulent kinetic energy, which fails to follow the experimental data in \fig\ref{fig:5c}.} 
\replace{The higher value of turbulent kinetic energy }{Higher $k$} is predicted near the wall in the proximity of dilution holes \replace{at positions $r/R_c = -0.75$ to $r/R_c = 0.75$}{for $-0.75\le r^\ast\le 0.75$}  and the lower \add{$k$} values \replace{at positions $r/R_c = -0.25$ to $r/R_c = 0.25$}{for $-0.25\le r^\ast\le 0.25$}\replace{. While experimental data suggest the maximum magnitude of turbulent kinetic energy at the centre of the plane, at positions $r/R_c = -0.25$ to $r/R_c = 0.25$}{, in contrast to the experimental data, where the maximum $k$ appears at the center of the plane for $-0.25\le r^\ast\le 0.25$}. 
\replace{A comparison of turbulent kinetic energy prediction with other turbulence models indicates maximum turbulence is predicted using the realizable $k-\epsilon$ model. Relatively high overall turbulence is predicted using the realizable k-epsilon model because the model predicted a rapid diffusion of the vortex core halfway before the fluid reached the dilution holes. }{The turbulent kinetic energy prediction indicates maximum turbulence using the realizable $k-\epsilon$ model, compared to other turbulence models, due to the rapid diffusion of the vortex core halfway before the fluid reached the dilution holes predicted using the realizable k-epsilon model.} \replace{This intensified mixing and interaction with the surrounding fluid and hence enhanced turbulence.}{It intensifies mixing and interaction with the surrounding fluid, yielding enhanced turbulence. } Besides, the prediction of \replace{a higher value of}{higher} turbulent kinetic energy values \remove{predicted} near the wall \replace{at positions $r/R_c = -0.75$ to $r/R_c = 0.75$}{in the region $-0.75\le r^\ast\le 0.75$} can be due to additional air entering the combustor through dilution holes that can create flow gradients and vortical structures near the wall enhancing the turbulence energy in that region. 
\\
\replace{The shear stress prediction shown in {\fig\ref{fig:5d}} is also not in trend with the experimental values and is negative throughout the plane and very high overall values are predicted.}{The shear-stress prediction is also inconsistent with the experimental data; the values are negative and overall very high throughout the plane, as shown in \fig\ref{fig:5d} possibly due to an}\remove{The high negative shear stress predicted on the dilution hole plane can be attributed to the} enhancement of turbulence \replace{due to}{arising by the} vortex diffusion upstream\replace{ and}{,} complex flow interactions and vortical structures \replace{present due to }{resulting from the} air entering the combustor liner through dilution holes. 
The turbulent viscosity profile \replace{of}{obtained using} the realizable $k-\epsilon$ model in \fig\ref{fig:7} shows considerably high magnitudes \remove{predicted} with a nearly flat profile throughout the plane \replace{from position $r/R_c = -0.75$ to $r/R_c = 0.75$}{for $-0.75\le r^\ast\le 0.75$}. 
The high magnitude of turbulent viscosity is predicted in the region of high turbulence in the realizable $k-\epsilon$ model due to model formulation. 
The nearly flat profile of the eddy viscosity suggests that the model assumes a relatively constant turbulent viscosity throughout the region, possibly due to the isotropic assumption of eddy viscosity. 
\replace{The present simulation shows the model’s inability to simulate the anisotropic behaviour of confined swirling flows and complex interaction between the swirling flow, turbulence and energy transfer processes accurately as found in gas turbine combustors which have led to discrepancies between the predicted flows and observed flow characteristics.}{The above analysis has clearly shown the inability of the realizable $k-\epsilon$ model to accurately predict the anisotropic behavior of confined swirling flows and complex interaction between the swirling flow, turbulence, and energy transfer processes as found in gas turbine combustors.}
%--------------------------------------------------------------------------------
\subsubsection{Standard $k-\omega$ model  predictions}
\begin{figure}[!tb]
	\centering
	\subfigure[Standard $k-\epsilon$]{\includegraphics[width=0.48\linewidth]{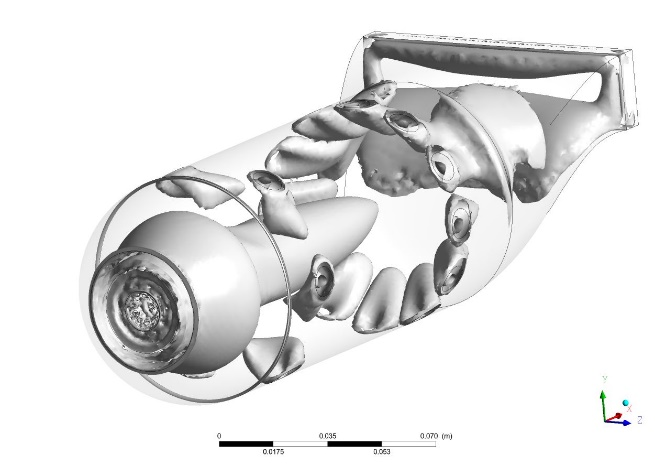}\label{fig:10a}}
	\subfigure[Realizable $k-\epsilon$]{\includegraphics[width=0.48\linewidth]{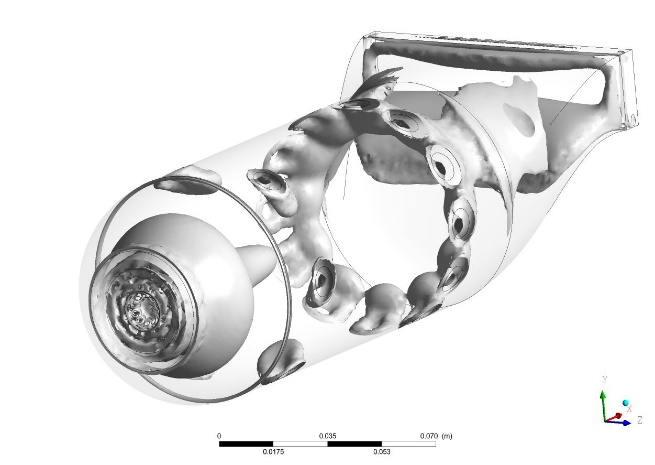}\label{fig:10b}}
	\subfigure[Standard $k-\omega$]{\includegraphics[width=0.48\linewidth]{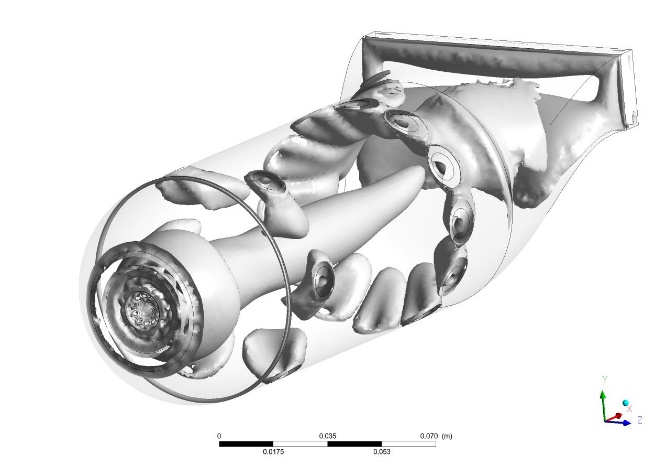}\label{fig:10c}}
	\subfigure[SST $k-\omega$]{\includegraphics[width=0.48\linewidth]{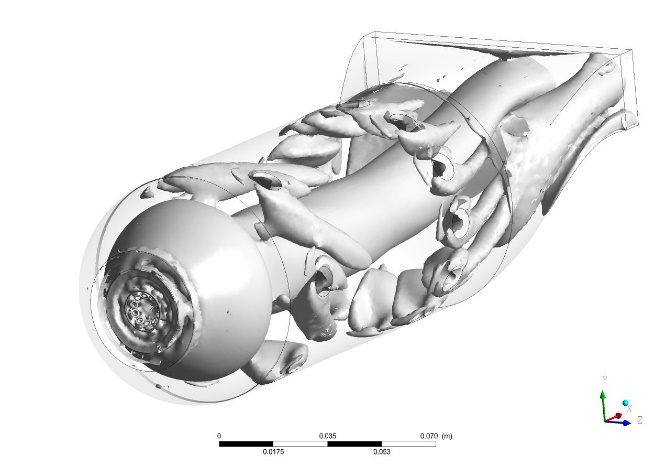}\label{fig:10d}}
	\subfigure[LPS-RSM]{\includegraphics[width=0.48\linewidth]{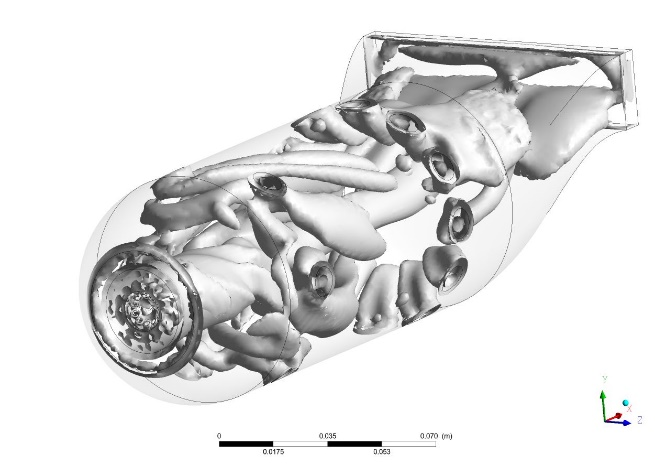}\label{fig:10e}}
	\caption{Comparison of Q-criterion predicted using various tutbulence models  (a) standard $k-\epsilon$, (b) realizable $k-\epsilon$, (c) Standard $k-\omega$, (d) SST $k-\omega$, (e) LPS-RSM.}\label{fig:10}
\end{figure}

%--------------------------------------------------------------------------------
%
\replace{In the axial velocity comparison between the predicted mean axial velocity using the standard $k-\omega$ model in {\fig\ref{fig:4a}}, it can be observed that the predicted mean axial velocity is inaccurately predicted and not in trend with the measured values.}{The axial velocity ($U/U_b$) predicted is also inaccurately predicted using the standard $k-\omega$ model and unmatched with experimental data, as shown in \fig\ref{fig:4a} at the primary holes plane ($z=50$ mm).} \replace{The predicted axial velocity again shows non-negative values indicating that the model has failed to predict the backflow present in the vortex flows.}{The predicted non-negatived axial velocity again indicates the failure of the model to predict the backflow present in the swirling vortex flows.}
\replace{This can again be attributed}{It is due} to the assumption of isotropic turbulence in the standard $k-\omega$ model\replace{ and the model formulation  which may not}{,  which imposes an inability to} fully capture the complex flow dynamics and mechanisms driving backflows in the vortex near primary holes of the combustor. 
\replace{T}{On the other hand, t}he predicted mean transverse \add{velocity} in \fig\ref{fig:4b} is in \replace {trend with experimental values and is in fairly}{reasonably} good agreement with the \replace{measured}{experimental} values. The position \add{($r^\ast = 0$)} where the transverse velocity switches from positive to negative\remove{, $r/R_c = 0.0$} is also accurately predicted. 
The \remove{predicted} turbulent kinetic energy is over-predicted \replace {and is not in agreement with}{to} the experimental values\add{at the primary holes,} as seen in \fig\ref{fig:4c}. The maximum values \remove{are} predicted near \replace{wall the}{the wall} due to the swirling motion \remove{and} are further enhanced by air entering from primary holes\replace{, whilst}{. In contrast,} the minimum values are predicted at the center at the vortex core, where experimental values indicate maximum turbulence. 
\replace{This discrepancy as aforementioned can be attributed to the incapabilities of}{Similar to the discrepancies in the axial velocity, } the standard $k-\omega$ model \add{has shown incapability} to adequately capture the unique flow characteristic of flows in can combustors (confined swirling flows with additional primary and dilution jets), especially the interaction between the vortex core and primary jets.
\\
\replace{The predicted magnitude of shear stress can also be said to be enhanced by the primary jets entering the combustor liner through primary holes. The experimental values also indicate the maximum same pattern, the predicted values are larger than experimental values and not aligned with them at both ends, positions $r/R_c = -0.75$ and $r/R_c = 0.75$. However, negligible shear stress is predicted at the centre position $r/R_c = 0.0$ or at the vortex core where maximum values of shear stress were experimentally recorded due to the development of }{Further, the predicted shear stress magnitude is enhanced by the primary jets entering the combustor liner through primary holes, as seen in \fig\ref{fig:4d}. While the maximum shear stress obtained using the standard $k-\omega$ model is similar to experimental data, other predicted shear stress values are larger than the experimental values and do not align at both ends ($r^\ast = \pm 0.75$). Furthermore, negligible shear stress is predicted at the center ($r^\ast = 0$) or at the vortex core where experimental values are maximum due to the development of the} backflow in the vortex core. The prediction of zero shear stress at the vortex core could be a limitation of the model to capture the flow physics adequately in that region. 
\\
\replace{On analysing the turbulence viscosity curve in {\fig\ref{fig:6}} it can be said that the maximum turbulent viscosity is predicted near the combustor liner walls at the edge of the vortex due to intense turbulent activity and shear effects in this region, further enhanced by air entering the combustor liner through primary holes.}{The maximum turbulent viscosity is predicted in \fig\ref{fig:6} near the combustor liner walls at the edge of the vortex core due to intense turbulence and shear effects in the region, further enhanced by air entering the combustor liner through primary holes.} 
\replace{The turbulence viscosity represents the level of turbulence-induced mixing and momentum transfer in the flow, at the edge of the vortex, there are significant velocity gradients and strong vorticity, which results in enhanced turbulence and mixing.}{The turbulence viscosity indicates the level of turbulence-induced mixing and momentum transfer in the flow; significant velocity gradients and vorticity results in enhanced turbulence at the edge of the vortex.} 
\replace{These conditions contributed to the generation of stronger turbulent eddies, contributing to the computation of higher turbulent viscosity in this region. On the other hand, with the flat profile of the turbulent viscosity at the centre of the vortex position $r/R_c = 0.0$ the computed velocity gradients and turbulence kinetic energy is lower, which has led to the flat profile.}{These conditions generate stronger turbulent eddies, contributing to the higher turbulent viscosity in this region. On the other hand, a flat turbulent viscosity profile at the center of the vortex position ($r^\ast = 0$) appears due to the lower velocity gradients and turbulence kinetic energy.}    
\\
\replace{On observing the standard $k-\omega$ model predictions on the dilution holes planes, the predicted axial velocity profile shown in {\fig\ref{fig:5a}} is not in trend with the experimental values, the maximum values are predicted at radial positions $r/R_c = -0.25$ and $r/R_c = 0.25$, while a slight trough can be seen at the centre of the plane at radial position $r/R_c = 0.0$. The transverse velocity shown in {\fig\ref{fig:5b}} is also underpredicted indicating model predicts diffusion of the vortex at the position which originated upstream at the primary zone by swirler.}{Further, the prediction using the standard $k-\omega$ model at the dilution holes plane ($z=130$ mm) for the axial and mean transverse velocities also deviates from the experimental data, as shown in \figs\ref{fig:5a} and \ref{fig:5b}. The maximum axial velocity is obtained at $r^\ast = \pm0.25$ with a slight trough at the center of the plane at $r^\ast=0$. The overpredicted transverse velocity indicates vortex diffusion at this plane, which originated upstream at the primary zone by swirler.} 
The diffusion of the vortex is also reflected in velocity vectors in \fig\ref{fig:8c}, streamlines in \fig\ref{fig:9c}, and $Q-$criterion in \fig\ref{fig:10c} as the fluid reaches the dilution holes plane. 
 \\
The turbulent kinetic energy ($k$) is overpredicted\add{,} as seen in \fig\ref{fig:5c}. The maximum values are obtained at the edge of the vortex close to the wall at \replace{radial positions $r/R_c = -0.75$ and $r/R_c = 0.75$}{$r^\ast = \pm 0.75$}, where velocity gradients are expected to be maximum, further enhanced by intensified mixing predicted between the vortex and the surrounding fluid due to vortex diffusion and additional air entering the combustor through dilution holes.  The minimum values are predicted at the core of the vortex at \replace{radial positions $r/R_c = 0.0$}{$r^\ast = 0$}, which contrasts with the experimentally maximum $k$ is recorded at the position. 
\\
\replace{The maximum shear stress is also predicted at the edge of the vortex close to the wall at radial positions $r/R_c = -0.75$ and $r/R_c = 0.75$  where velocity gradients are expected to be high as seen in {\fig\ref{fig:5d}}.}{The maximum shear stress predicted at the edge of the vortex close to the wall at $r^\ast=\pm 0.75$ as compared with the maximum recorded experimentally at the center ($r^\ast=0$), as seen in \fig\ref{fig:5d}, which presumably is at the center of the vortex core characterized by high rotational velocities and strong velocity gradients.} When the vortex diffuses, it undergoes deformation and spreads out in the flow field\replace{. This diffusion process can create}{, thus creating} regions of high-velocity gradients and enhanced shear stress at the edge of the vortex.  \replace{The interaction between the turbulent flow, vortex and dilution jets, can lead to increased turbulent transport and mixing, resulting in elevated shear stress levels. However, the maximum shear stress is experimentally recorded at the centre position $r/R_c = 0.0$, presumably at the centre of the vortex core. The vortex core is characterized by high rotational velocities and strong velocity gradients. Interaction between the swirling flow and the surrounding liquid creates}{An interaction between the swirling turbulent flow, vortex and dilution jets, and the surrounding liquid can increase turbulent transport and mixing, thereby creating} significant shear stress within the vortex structure \citep{Dixon2014}. 
\\
The underprediction of turbulent kinetic energy and shear stress \remove{in this region} can be attributed to the inherent limitation of isotropic viscosity assumptions and simplifications in the standard $k-\omega$ model\remove{. The $k-\omega$ model is based on certain assumptions and empirical correlations} that may not \add{{accurately}} capture the complex flow phenomenon \remove{accurately} as in the present case.  
 \add{\fig\ref{fig:7} shows that} the maximum magnitude of the turbulent viscosity is predicted at the outer edge of the vortex core due to the complex flow behavior and velocity gradients in that region\remove{, shown in {\fig\ref{fig:7}}}. \remove{In swirling flows, such as those found in vortex cores, there is a strong rotation of fluid particles around the central axis. This rotation leads to the development of high-velocity gradients and vorticity.}
At the outer edge of the vortex,  \remove{where} the flow is rapidly rotating, \add{and} the velocity gradients are highest\add{,} which contributes to the generation of turbulence and mixing of fluid elements, leading to an increased turbulent viscosity.  \replace{As in}{In} the $k-\omega$ model, the specific dissipation rate ($\omega$) equation (\eqn\ref{eq:24}) includes a production term \remove{related to the mean flow strain rate. This term} represents the transfer of kinetic energy from the mean flow \add{strain rate} to the turbulence\replace{, and it}{and thus} influences the turbulent viscosity. In regions of high strain rate\remove{or strong mean flow gradients}, the production term  (See section \ref{sec3.2.3.2}) can lead to higher values of $\omega$ and, consequently, higher turbulent viscosity. However, in the core region of a vortex, the mean flow strain rate is typically lower compared to the outer regions\replace{. This is because}{, as} the flow within the core is primarily characterized by rotation rather than deformation. This lower strain rate can result in a reduced production term and\replace{, consequently, a lower predicted}{predict a lower} turbulent viscosity. 
\replace{However, readers should note that}{Notably, } turbulent viscosity is not a physical property but \remove{rather} a modeling concept introduced \add{by turbulence models } to represent the complex nature of turbulence \replace{It is typically introduced by turbulence models, such as RANS models, which provide closure equations to approximate the effects of turbulence on the flow}{by providing closure equations to approximate the effect of turbulence on the flow}. 
%
%\\
\subsubsection{SST $k-\omega$ model predictions}
\replace{The axial velocity predicted on the primary hole plane using the SST $k-\omega$ model is also not accurate and also not in trend with the experimental measurement as seen in {\fig\ref{fig:4a}}}{\fig\ref{fig:4a} depicts the axial velocity predicted on the primary hole plane ($z=50$ mm) using the SST $k-\omega$ model, which is inaccurate and inconsistent with the experimental measurements}. However, amongst the two-equation \add{turbulence} models used \add{in this work},  only the SST $k-\omega$ model has \replace{been able to predict}{predicted} the negative axial velocity \remove{values} in the vortex core region at \replace{radial positions $r/R_c = -0.25$ and $r/R_c = 0.25$}{$r^\ast=\pm 0.25$} indicating that the model has \replace{managed to capture}{captured} backflow found in complex vortex flows to some extent. The velocity vectors in \fig\ref{fig:8d} also reflect that the SST $k-\omega$ model predicts the presence of backflow in the vortex core. 
\\
The transverse velocity predictions are also inaccurate\add{,} and magnitudes are \replace{greater}{more significant} than the experimental measurements\remove{ of the same}, as \remove{can be} seen in \fig\ref{fig:4b}. \replace{The main reason for inaccuracy in the above prediction could be that SST $k-\omega$ is also based on isotropic turbulence. It is already established that turbulence in the confined swirling flow is anisotropic, leading to errors in the model prediction.}{As established in previous sections, turbulence in the confined swirling flow is anisotropic, leading to errors in the model prediction, as the SST $k-\omega$ model is based on isotropic turbulence.}  \add{Further,} the SST $k-\omega$ \add{model} is \remove{also} a linear eddy viscosity model, \replace{it assumes}{assuming} that turbulent viscosity is proportional to turbulent kinetic energy. This assumption \replace{is made to simplify}{simplifies} the mathematical formulation of the model. However, it may not be suitable for \replace{the accurate prediction of}{accurately predicting} confined swirling flows. 
\\
The predicted turbulence kinetic energy in \fig\ref{fig:4c} is also not in agreement with the experimental data\replace{ and}{;} like other assessed two-equation models\add{,} it \replace{is}{has} also \replace{predicted}{indicated} maximum at the edge of the predicted vortex core at \replace{radial positions $r/R_c = -0.75$ and $r/R_c = 0.75$}{$r^\ast =\pm 0.75$}, close to the walls, and minimum magnitude \remove{is predicted} at the vortex core at \replace{radial positions $r/R_c = -0.25$ and $r/R_c = 0.25$}{$r^\ast =\pm 0.25$}.  \add{Similarly, the maximum and minimum magnitudes of shear stress are predicted at $r^\ast =\pm 0.75$ and $r^\ast =\pm 0.25$, respectively, in \fig\ref{fig:4d}.} 
\replace{The maximum values occur at the vortex edge due to the most intense turbulence. The interaction of the primary jets with the surrounding fluid further enhances the turbulent intensity. It becomes strongest at the vortex, where shear forces and velocity gradients are most significant.}{The maximum magnitude of turbulent kinetic energy is predicted at the vortex edge because this is where the turbulence is most intense. The turbulence is also further enhanced by the interaction of the primary jets with the surrounding fluid and is strongest at the vortex where shear forces and velocity gradients are greatest.}  
\remove{The maximum magnitude of shear stress is also predicted in the same region at radial positions $r/R_c = -0.75$ and $r/R_c = 0.75$, and the minimum is predicted around the vortex core at radial positions $r/R_c = -0.25$ and $r/R_c = 0.25$. 
The lower level of turbulence predicted at the vortex core can be attributed to the model’s inherent incapabilities to accurately simulate confined swirling flows due to the model’s assumption on which its formulation is based and the assumption of isotropic turbulence due to which the model may not fully capture the complexities of vortex flows in such confined environment.} 
\\
\replace{On examining the predicted turbulent profile of the SST $k-\omega$ model in {\fig\ref{fig:6}}, it can be seen that the nearly flat profile predicted across the plane obtained indicates almost constant eddy viscosity.}{The turbulent viscosity ($\mu_t$) profile predicted using the SST $k-\omega$ model in \fig\ref{fig:6} is nearly flat across the plane, indicating almost constant eddy viscosity.} The predicted \remove{turbulent viscosity} profile is similar to the standard $k-\epsilon$ model\replace{, however, is}{but} much lower in magnitude. 
\replace{This could be the reason why }{It must have resulted in} the vortex core \replace{is sustained}{continuing} till the combustor exit\add{,} as seen in the $Q-$criterion in \fig\ref{fig:10d} and velocity streamlines in \fig\ref{fig:9d}.
In the case of the SST $k-\omega$ model, the predicted lower $\mu_t$ may contribute to the sustained vortex structure. As the fluid moves within the vortex core, the lower $\mu_t$ facilitates the interaction between fluid layers with different velocities and vorticity, leading to a more coherent and stable vortex structure. On the other hand, in the standard $k-\epsilon$ model, where higher $\mu_t$ values are predicted, particularly at the core region, the vortex may be more prone to diffusion and dissipation. Nevertheless, both models failed to fully capture the anisotropic nature of the turbulence at the vortex core region. 
\\
\replace{When comparing the predicted mean axial velocity using the SST $k-\omega$ model with the experimental measurements at the dilution holes plane in {\fig\ref{fig:5a}}, it becomes evident that the mean axial velocity, like other evaluated two-equation models, does not exhibit a trend consistent with the experimental values.}{At the dilution holes plane ($z=130$ mm), the mean axial velocity predicted using the SST $k-\omega$ model does not exhibit a trend consistent with the experimental values in \fig\ref{fig:5a}.} It is, however, noteworthy that the predicted mean axial velocity, although overpredicted, is relatively closer to the experimental values compared to the other evaluated two-equation models. 
\\
\replace{The predicted mean transverse velocity shown in {\fig\ref{fig:5b}}, while still inaccurate, shows better agreement with other assessed two-equation turbulence models. The higher magnitude of the transverse velocity, exceeding that of other evaluated two-equation models, suggests a stronger swirl component and supports the observation of a sustained vortex core as discussed earlier. The above-mentioned disparities in predicted mean axial and transverse velocity with other assessed two-equation turbulence models could be attributed to the difference in the formulation of turbulent viscosity in the SST $k-\omega$ model. The SST $k-\omega$ model employs a different approach to estimate turbulent viscosity, taking into account the effects of the turbulent kinetic energy and the specific dissipation rate and additional formulation. This formulation may lead to variations in the predicted flow characteristics when compared to other turbulence models such as other evaluated two-equation models (see {\eqn\ref{eq:38}}).} {The predicted mean transverse velocity in \fig\ref{fig:5b} shows inaccurate but better agreement with experimental data. The higher magnitude of the transverse velocity, exceeding that of other evaluated two-equation models, suggests a stronger swirl component and supports the observation of a sustained vortex core, as discussed earlier. These disparities in predicted mean axial and transverse velocity, with other assessed two-equation turbulence models, could be attributed to the difference in the formulation of turbulent viscosity in the SST $k-\omega$ model. The SST $k-\omega$ model employs a different approach to estimate turbulent viscosity, taking into account the effects of the turbulent kinetic energy and the specific dissipation rate and additional formulation. This formulation may lead to variations in the predicted flow characteristics when compared to other turbulence models, such as other evaluated two-equation models (see \eqn\ref{eq:38}). }
\\
\replace{When observing the turbulent kinetic energy on the dilution hole plane in {\fig\ref{fig:5c}}, it can be seen that lower kinetic energy is predicted throughout the plane than using the other two-equation models. Maximum values are still predicted at the vortex edge on the dilution holes plane at radial positions $r/R_c = -0.75$ and $r/R_c = 0.75$ using the SST $k-\omega$ model as a result of the complex flow dynamics in that region and minimum values are predicted at the centre of the dilution holes plane from radial positions $r/R_c = -0.25$ to $r/R_c = 0.25$.}{The turbulent kinetic energy is underpredicted  in \fig\ref{fig:5c} using the SST $k-\omega$ model, compared to other two-equation models, throughout the dilution hole plane ($z=130$ mm) with a nearly flat profile for $-0.5\le r^\ast \le 0.5$, maximum values at the vortex edge ($r^\ast = \pm 0.75$), and minimum values in the mid ($-0.25\le r^\ast \le 0.25$) of the plane.} \remove{As aforementioned, the vortex core edge is characterized by strong velocity gradients and intense turbulence due to the swirling motion of the flow. These factors contribute to the higher turbulent kinetic energy levels near the vortex core edge and are further enhanced by the interaction of dilution jets with the convecting fluid which leads to an increase in mixing and turbulence level. However, at most of the plain, at the positions $r/R_c = -0.5$ to $r/R_c = 0.5$, a nearly flat profile of turbulent kinetic energy is predicted,} \add{Further,} the model failed to predict the peak at the center of the plane or \remove{the center of} the vortex core, as observed in the experimental measurement. 
These observed discrepancies can be due to \replace{the model’s}{an} inherent inability of \add{the model} to \remove{accurately} simulate confined swirling flows \add{accurately} owing to model formulations including isotropic turbulence assumption. 
\replace{On observing the predicted shear stress in {\fig\ref{fig:5d}}, the maximum shear stress is predicted at the vortex core edge due to the aforementioned intense velocity gradients and flow deformation in this region further enhanced by the interaction of dilution jets with swirling flows. The model has predicted lower shear stress at the centre position $r/R_c = -0.5$ to $r/R_c = 0.5$, the reasons for these discrepancies are mentioned above.}{Furthermore, the lower shear stress is obtained for $-0.5\le r^\ast \le 0.5$, and maximum shear stress at the dilution holes plane in \fig\ref{fig:5d} is predicted using the SST $k-\omega$ model at the vortex core edge due to the intense velocity gradients and flow deformation, enhanced by the interaction of dilution jets with swirling flows in this region.} 
\\
The turbulent viscosity is computed as maximum around the center of the vortex core \add{where the turbulent kinetic energy is lowest} and minimum at the vortex edge, shown in \fig\ref{fig:7}\replace{. The turbulent viscosity is a measure of the resistance of the fluid to shearing. It is calculated from the turbulent kinetic energy and the specific dissipation rate in}{using} the  SST $k-\omega$ model. \replace{The turbulent viscosity is calculated highest at the centre of the vortex core where the turbulent kinetic energy is computed lowest. This is because the turbulence is being dissipated and the fluid is less able to resist shearing.}{It results from less shearing resistance by the fluid in the presence of turbulent dissipation.}
\replace{On comparing the computed turbulent viscosity using the SST $k-\omega$ model with a computed turbulent velocity of other two-equation models in {\fig\ref{fig:7}}, it can be seen}{Further, the magnitude of the} turbulent viscosity \remove{magnitude} computed at the dilution hole using the SST $k-\omega$ model is lower than the other assessed two-equation models. \replace{This discrepancy can be attributed to the different formulations and closure assumptions employed in each model.}{This discrepancy is due to varied formulations and closure assumptions employed in the turbulent models.}  The SST $k-\omega$ model combines \replace{elements of the}{the standard} $k-\epsilon$ and \add{the standard} $k-\omega$ models \remove{and aims to provide improved}{to improve} predictions in \replace{a wider }{the broader}range of flow conditions. 
However, the specific coefficients, turbulence closure assumptions, and treatment of turbulence production, dissipation, and length scales in the SST $k-\omega$ model can lead to variations in turbulent viscosity predictions. \replace{Still, from the assessment it can be stated}{Nevertheless, the} SST $k-\omega$ model demonstrated superior performance compared to the other evaluated two-equation models in the present flow conditions.  
%\\
\subsubsection{Linear Pressure Strain - Reynolds Stress Model (LPS-RSM) predictions}
\replace{On observing the predictions using the LPS-RSM (Linear Pressure Strain Reynolds Stress Model), it can be seen in {\fig\ref{fig:4a}} that the mean axial velocity is overestimated, however, is in trend with the measured values.}{The mean axial velocity obtained using LPS-RSM is overestimated but consistent with the experimental results in \fig\ref{fig:4a}.} The model has \replace{managed to capture}{accurately captured} the negative values \add{of the axial velocity} in the highly turbulent vortex core region. In \replace{highly turbulent vortex core}{these} regions, as \replace{experimental data of the reference study suggest}{experimentally observed \citep{Heitor1986,Heitor1985}}, the axial velocity typically becomes negative\replace{. This is because as the vortex core is rotating, the fluid is being forced to move in a circular motion. This results in}{as the rotating vortex core forces the fluid to move circularly, developing} complex velocity gradients and vorticity patterns.
\replace{The LPS-RSM managed to capture the negative axial velocity in a highly turbulent vortex core region because it takes into account the anisotropy of turbulence.}{LPS-RSM considers the anisotropy of turbulence and, thus, has successfully captured the negative axial velocity in the highly turbulent core regions; in contrast, the two-equations ($k-\epsilon$, and $k-\omega$) turbulence models failed due to their inherent isotropic assumptions.}
\replace{In the LPS-RSM, the}{The} pressure strain term \add{($\phi_{ij}$) in LPS-RSM} is decomposed into slow and rapid pressure strain terms along with the wall reflection term (\eqn\ref{eq:50})\replace{. The}{, wherein the} slow pressure strain term ($\phi_{ij,1}$, \eqn\ref{eq:52}) contains the Reynolds stress anisotropic tensor \replace{({\eqn\ref{eq:52}})}{($b_{ij}$)}, which is a measure of anisotropy. The term is proportional to the amount of anisotropy in the Reynolds stress tensor\replace{. The slow pressure strain tensor can be thought of as a}{ and, thus, can} measure \remove{of} the tendency of the Reynolds \replace{S}{s}tress \replace{T}{t}ensor to relax back to isotropy.  
The anisotropic stress tensor \replace{is modelled}{modeled} as a linear function of the turbulent kinetic energy ($k$)\replace{. This means that}{yields} the slow pressure strain term \replace{is proportional to the amount of turbulent kinetic energy in}{proportional to $k$ of} the flow.
\replace{As the turbulent kinetic energy increases, the slow pressure strain term also increases, which helps to promote}{Increasing $k$ manifests the slow pressure strain, which promotes} isotropy in a turbulent flow\replace{. The}{, as the} slow pressure strain term \replace{works by transferring}{transfers} energy from the anisotropic eddies to the isotropic eddies through the action of pressure forces.
\replace{This helps to restore the isotropy of the turbulent flow and reduces the problems associated with anisotropy. The pressure forces act on the turbulent eddies, they tend to equalize the stresses in the turbulent flow. Hence, pressure forces tend to shrink the anisotropic eddies.}{It helps to restore the isotropy of the turbulent flow and, thus, reduces the problems associated with anisotropy as the pressure forces acting on the turbulent eddies tend to equalize the stresses in the turbulent flow and shrink the anisotropic eddies.} 
The rapid pressure strain term ($\phi_{ij,2}$, \eqn\ref{eq:53}) \replace{is a measure of}{measures} the interaction between the pressure and the velocity gradients\replace{. The rapid pressure strain term}{ and} is responsible for the rapid distortion of turbulence\remove{,} by transferring energy from the mean flow to the Reynolds stresses. \replace{It}{Despite being a symmetric tensor, it } is an anisotropic term\remove{, despite being a symmetric tensor,} since the different components of the pressure-strain tensor \replace{are not equal}{unequal}. 
In particular, the rapid pressure-strain \remove{term} is \replace{larger}{more significant} when the mean strain rate tensor \remove{term} ($\overbar{S}_{ij}$, \eqn\ref{eq:3}) is aligned with \remove{the direction of} the Reynolds stresses. \remove{It can do this more effectively when the mean stress rate is aligned with the direction of the Reynold stress.} 
The LPS-RSM often overestimates the velocity \replace{which is mainly due to the LPS-RSM assuming that the pressure-strain correlation is linear in the Reynolds stresses which means that the pressure strain correlation is proportional to}{mainly due to the linear relationship between the pressure-strain and} the root mean square (RMS) of the Reynolds stresses.
\\
\replace{The pressure strain correlation is a complex phenomenon, and it is not fully understood, however, it plays an important part and turbulence modelling. It is mostly understood as a term in the Reynolds stress transport equation that describes how the pressure field interacts with the Reynolds stresses.}{The pressure strain correlation is a complex and partially understood phenomenon but plays a vital role in turbulence modeling. It is expressed by the Reynolds stress transport equation describing the pressure field interaction with the Reynolds stresses.}
The pressure-strain correlation can be quadratic or \remove{even} higher order in complex flows.  
\replace{The Linear Pressure correlation aligns the pressure field with the strain rate, this alignment creates a coupling between the pressure field and the velocity field which increases the momentum transfer between different layers of fluid.}{The Linear pressure correlation aligns the pressure field with the strain rate and couples the pressure and velocity fields, which increases the momentum transfer between different layers of fluid.}
\replace{This causes an increase in}{It enhances the} turbulent viscosity and can \replace{also lead to an overestimation of}{overestimate} the velocity of the fluid \citep{Pope2000,Biswas2002,Kajishima2017,Bernard2018}.
\\
\replace{Hence, due to the above-mentioned reasons the transverse velocity, turbulent kinetic energy, and shear stress in the primary hole plane are overestimated (see {\figs\ref{fig:4b}}-{\ref{fig:4d}})  are overestimated on the primary hole plane. However, these results are in trend with the experimental data.} {The above discussion thus justifies the overestimation, but consistent trends with the experimental data, of the transverse velocity, turbulent kinetic energy, and shear stress at the primary hole plane by using the LPS-RSM in \figs\ref{fig:4b} -- \ref{fig:4d}.}
The predicted turbulent viscosity profile using the LPS-RSM in \fig\ref{fig:6} is similar to the predicted turbulent kinetic energy in \fig\ref{fig:4c} because the turbulent viscosity is proportional to the square of the turbulent kinetic energy (see \eqn\ref{eq:10}), in LPS-RSM, similar to the standard $k-\epsilon$ model.
\replace{Due to consideration of anisotropy and model formulation, the model predicted}{Furthermore, anisotropic consideration in the LPS-RSM predicts} the swirling more accurately than the two-equation models, \replace{hence, it can be}{as} seen in \figs\ref{fig:8e}, \ref{fig:9e} and \ref{fig:10e}  that \remove{the model managed to preserve} the swirling vortex core \add{is preserved} till the combustor exit \add{comparatively } better than the two-equation counterparts.
\\
\replace{The predicted axial velocity using LPS-RSM at the dilution holes plane shows an undulated profile in {\fig\ref{fig:5a}}, the axial velocity is overpredicted, however, the waviness predicted shows that predictions are more in trend with the experimental data than the other assessed two-equation models. This is likely because the LPS-RSM model takes into account anisotropy as aforementioned. }{Similarly, the axial velocity at the dilution holes plane ($z=130$ mm) predicted using LPS-RSM is undulated and overpredicted in \fig\ref{fig:5a}; however, waviness indicates better consistency with experimental data than other assessed two-equations models likely because LPS-RSM model accounts for anisotropy, as discussed before.} 
The transverse velocity is also more accurately predicted using the LPS-RSM than other assessed two-equation models as shown in \fig\ref{fig:5b}\replace{, the predicted transverse velocity profile indicates the vortex core as it switches from positive to negative at almost the same position and is similar magnitude overall}{. The vortex core, as the transverse velocity switches from positive to negative, is indicated closely at the same position, and the overall magnitude is also similar} to the experimental data. 
\\
{The turbulent kinetic energy is underpredicted but overall more accurate using the LPS-RSM model than two-equation models, however, is predicted flat at the centre of the plane.}{The turbulent kinetic energy is underpredicted with a flat profile at the center of the dilution holes plane using LPS-RSM in \fig\ref{fig:5c}. However, it is overall more accurate than two-equation models.} 
\replace{However, the experimental data and LPS-RSM prediction of}{Both experimental and LPS-RSM predictions for the} transverse velocity show that the center of the vortex core lies on the \add{center of the } dilution holes plane\remove{centre}. 
\replace{The centre of the vortex is a more chaotic region of the flow, with strong vortices and significant fluctuations which is indicated by the experimental data of turbulent kinetic energy in {\fig\ref{fig:5c}} as it shows a peak on this position.} {The center of the vortex is highly chaotic, with strong vortices and significant fluctuations, and indicated by a peak in the experimental data of turbulent kinetic energy in \fig\ref{fig:5c}.} 
Hence, \remove{it can be stated that} the linear pressure strain correlation in the LPS-RSM model\replace{, model} has underestimated the \add{energy} transfer \remove{of energy} between different components of Reynolds stresses, \replace{hence, }{due to which} the model has failed to predict turbulence \remove{in}accurately at the vortex core on the dilution holes plane.
\replace{On observing the predicted shear stress on the dilution hole plane, the predicted profile}{The shear stress profile on the dilution holes plane} obtained using LPS-RSM is also flat and shows negligible shear stress in \fig\ref{fig:5d}\replace{, the reasons for which are mentioned above}{due to the abovementioned reasons}. 
{The predicted turbulent viscosity has undulated profile and is relatively high compared to the other two-equation models except for Realizable $k-\epsilon$.}{Similar to the axial velocity profile in \fig\ref{fig:5a}, the turbulent viscosity profile in \fig\ref{fig:7} is undulated and relatively higher compared to the other two-equation, except for realizable $k-\epsilon$, models.}  The relatively higher turbulent viscosity indicates a prediction of a lower turbulence dissipation rate\remove{ at this position}. 
\\
\replace{Despite these shortcomings, the authors would like to state that the predictions using  the}{Despite the abovementioned shortcomings,} LPS-RSM \add{ predictions} are still more accurate than the assessed two-equation models. The overall findings can be summarized as follows:
\begin{itemize}
	\item 
	\replace{The isotropic turbulence assumption of the standard $k-\epsilon$ model, which assumed equal turbulence in all directions at a given point, fails to accurately represent the turbulent transport processes and vortical structures.}{The standard $k-\epsilon$ model, assuming the isotropic turbulence, i.e., uniform turbulence in all directions at a given point,  inaccurately represents the turbulent transport processes and vortical structures for the confined swirling flows.} 
	\replace{As a result, the model does not accurately capture the flow characteristics and complexities of confined swirling flows. Hence, axial velocity, transverse velocity, turbulence kinetic energy and shear stress are inaccurately predicted at both primary hole and dilution hole planes. The model failed to capture the flow reversal at the centre of the vortex and secondary flow features.}{The model, thus, inaccurately captures the flow characteristics and complexities of confined swirling flows, such as axial velocity, transverse velocity, turbulence kinetic energy, and shear stress at both primary hole and dilution hole planes. Further, the model has failed to capture the flow reversal at the center of the vortex and secondary flow features. }
	\replace{The model also failed to capture the concentration and intensification of turbulence due to the swirling motion and flow recirculation due to which aforementioned turbulence kinetic energy is inaccurately predicted. The $q-$criterion and the underpredicted velocity profile at the dilution hole clearly showed that the vortex core progressively dilutes as the flow convects downstream towards the dilution holes, which is not in agreement with the experimental observations. }{Broadly, the model predictions are inconsistent with the experimental observations.}
	\item 
	\replace{The axial velocity predicted using the realizable $k-\epsilon$ model was overpredicted and not in agreement with experimental values at both primary and dilution holes planes, and also was non-negative at the centre of the primary hole plane, similar to the standard k-epsilon model, indicating the model's inability to capture the central backflow as observed in the experimental data. The transverse velocity is also highly underpredicted at both positions, indicating a very weak swirl immediate diffusion of the vortex core as it begins to form compared to the dominant swirl recorded in the experimental data. The turbulence kinetic energy was overpredicted, with the maximum magnitude at the edge of the vortex and near the wall and the minimum at the centre of both planes. This trend was predicted because the model was simulating vortex diffusion which was also confirmed by the $q-$criterion which enhanced the turbulence due to intensified mixing between the vortex core and surrounding fluid, which also lead to inaccurate values of shear stress.}{Similar to the standard $k-\epsilon$ model, the realizable $k-\epsilon$ model failed to capture the flow phenomena accurately compared to the experimental observations. While the axial velocity is overpredicted, the transverse velocity is highly underpredicted at both primary and dilution hole planes. The non-negative axial velocity at the center of the primary hole plane indicates the inability of the model to capture the central backflow observed in experiments. The highly underpredicted transverse velocity indicates a weak swirl and instant diffusion of the vortex core as it begins to form compared to the dominant swirl recorded in the experimental data. The turbulence kinetic energy is also overpredicted, with the maximum magnitude at the edge of the vortex and near the wall and the minimum at the center of both planes. This trend appeared because the model simulates vortex diffusion, as confirmed by the $Q-$criterion, which enhanced the turbulence due to intensified mixing between the vortex core and surrounding fluid. It further leads to inaccurate shear stress.}
	\replace{The realizable k-epsilon turbulence is a modified form of the standard $k-\epsilon$ model, changes have been made to ensure that the model remains physical such as preserving the positivity of normal stress and maintaining the Schwarz inequality. However, the model has not been attuned to confined swirling flows, leading to less accurate predictions than the standard model. The model also incorporates additional terms and coefficients to improve turbulence prediction, but its isotropic formulation and lack of calibration limit its accuracy for confined swirling flows.}{While the realizable $k-\epsilon$ model preserves the positivity of normal stress, maintains the Schwarz inequality, and incorporates additional terms and coefficients to improve turbulence prediction, it has not been attuned to confined swirling flows due to an isotropic formulation and lack of calibration limit, thereby leading to less accurate predictions than the standard $k-\epsilon$ model.}
	\item 
	\replace{The standard $k-\omega$ model failed to accurately predict axial velocity profiles and backflow in the vortex at the primary hole plane, while the transverse velocity was captured reasonably well, while the axial velocity was captured inaccurately and overpredicted at the dilution hole plane and transverse velocity was underpredicted particularly at the dilation holes plane, indicating vortex diffusion at this position, which was not observed in the experimental measurement. The predicted turbulent kinetic energy is overpredicted, with maximum values near the vortex edge close to the wall. This can be explained due to predicted intensified mixing between the vortex and the surrounding fluid due to vortex diffusion which was also noticed in $q-$criterion and additional air entering the combustor through dilution holes. On the contrary, the minimum values of turbulent kinetic energy are predicted at the vortex core, opposing experimental observations. The model also predicted maximum shear stress at the edge of the vortex close to the wall, consistent with high-velocity gradients in that region. However, the predicted shear stress at the vortex core, where maximum values were experimentally recorded due to backflow development, was negligible at both primary and dilution hole planes.}{The standard $k-\omega$ model predictions have also shown inconsistency with the experimental results. The axial velocity profiles and backflow in the vortex are inaccurate, whereas transverse velocity is captured reasonably well at the primary hole plane. The axial velocity is inaccurate and overpredicted at the dilution hole plane, and the transverse velocity is underpredicted. The vortex diffusion observed, not in experiments, suggests additional air entry to the combustor through dilution holes, as noticed in $q-$criterion. Like the $k-\epsilon$ model,  the turbulent kinetic energy is overpredicted with maximum values near the vortex edge close to the wall because of intensified mixing between the vortex and the surrounding fluid due to vortex diffusion. Contrastingly, the minimum values of turbulent kinetic energy are predicted at the vortex core, as opposed to experimental observations. While the maximum shear stress consistent with high-velocity gradients is predicted at the edge of the vortex close to the wall, experiments recorded maximum values at the vortex core due to backflow development, which is negligible at both primary and dilution hole planes.} 
	\replace{The isotropic turbulent viscosity assumptions and lack of a mechanism in the model to capture the intricacies of vortex flow and its interaction with surrounding fluid in confined swirling flows caused such discrepancies.}{The reasons for discrepancies in the standard $k-\omega$ model predictions are similar to those with $k-\epsilon$ models, i.e.,  accounting for isotropic turbulence and an absence of incorporating vortex/swirling flows.}
	\item 
	\replace{The SST $k-\omega$ model does not accurately predict the axial velocity profiles on the primary hole plane, but it successfully captures the presence of backflow within the vortex core region. The transverse velocity predictions are also inaccurate but exhibit better agreement with other two-equation models. The axial velocity and transverse velocity predictions were inconsistent with the experimental measurement on the dilution hole plane, however, were better among the assessed two-equation models. The transverse velocity predictions and $q-$criterion indicated that the vortex core remained at the dilution holes plane and continued to the combustor exit as the flow convected towards it, which was consistent with the experimental observations. The model overpredicts turbulent kinetic energy, particularly at the vortex edge, and underpredicts it at the vortex core, opposing experimental observations at both positions. The predicted shear stress is highest at the vortex core edge due to intense velocity gradients, while negligible shear stress is predicted at the vortex core itself, which is again inconsistent with the experimental data at both primary and dilution hole planes. }{The SST $k-\omega$ model predictions are inaccurate but exhibit better agreement with experimental results than the other two-equation models; it has accurately captured the presence of backflow within the vortex core region in the axial velocity profiles on the primary hole plane. However, axial and transverse velocity predictions are inconsistent but better among the assessed two-equation models against the experimental measurement at the dilution hole plane. Further, the transverse velocity and $Q-$criterion indicated that the vortex core remained at the dilution holes plane and continued to the combustor exit as the flow convected towards it, consistent with the experimental observations. The model overpredicts turbulent kinetic energy, particularly at the vortex edge, and underpredicts it at the vortex core, opposing experimental observations at both positions. The predicted shear stress is highest at the vortex core edge due to intense velocity gradients and negligible shear stress at the vortex core itself. Again, it is inconsistent with the experimental data at both primary and dilution hole planes. }
	\replace{The SST $k-\omega$ model also assumes turbulence is isotropic, and the turbulent confined swirling flow is highly anisotropic, hence, discrepancies with experimental measurements are predicted. The SST $k-\omega$ model employs a different approach to estimate turbulent viscosity, taking into account the effects of the turbulent kinetic energy and the specific dissipation rate with the additional formulation. This may have led to variations in the predicted flow characteristics when compared to other turbulence models such as other evaluated two-equation models.}{While the SST $k-\omega$ model also assumes isotropic turbulence, it employs a different approach to estimate turbulent viscosity by accounting for turbulent kinetic energy and the specific dissipation rate. It has varied and improved predicted flow characteristics compared to other evaluated two-equation turbulence models.}
	\item 
	\replace{The LPS-RSM model overpredicted the mean axial velocity at the primary hole plane but was able to capture the negative axial velocity in the vortex core region, which is in agreement with experimental data. The model also overestimated the transverse velocity, turbulent kinetic energy, and shear stress on the primary hole plane. However, these results were in trend with the experimental data. This could be attributed to the model's consideration of turbulence anisotropy using the slow and rapid pressure strain terms in the pressure strain train terms The overestimation of velocity and turbulence in the predictions can be attributed to the model's assumption of a linear pressure strain correlation which can be quadratic or even of higher order in complex flows. The mean axial velocity on the dilution holes was overestimated and exhibited an undulated profile. However, it was more in line with the experimental measurement than the other two-equation models. The transverse velocity was also more accurately predicted than the other assessed two-equation models. The turbulent kinetic energy was underpredicted but was also more accurately predicted than the two-equation models, however, inaccuracies were noticed particularly at the dilution holes plane where the model failed to capture the peak at the centre of the plane as found in the experimental measurement. The shear stress was predicted fairly accurately on the primary holes plane, however, on the dilution holes plane was underpredicted, the model failed to predict the peak at the centre of the plane as found in experimental data. Nevertheless, the q-criterion showed that the model has managed to preserve the swirling vortex core till the combustor exit as specified in experimental measurements.  The noticed discrepancies between the experimental measurement and simulation results can be attributed to the Linear Pressure Strain Correlation in LPS-RSM, due to which the model does take into account complexities of turbulence such as non-linear interaction between velocity gradients and hence the model has not been able to accurately simulate complexities related to swirling flows, vortex core and its interaction with the surrounding fluid. Nonetheless, as it accounts for anisotropy in turbulence, it still performed better than the two-equation models.}{The LPS-RSM predictions consistently agree with the experimental measurement and are more accurate than the other two-equation models for the turbulent swirling flow characteristics on the primary and dilution hole planes. The mean axial and transverse velocity, turbulent kinetic energy, and shear stress are overestimated at the primary hole plane. The negative axial velocity in the vortex core region agrees well with experimental data. It is due to anisotropic turbulence consideration by splitting the pressure strain train into slow and rapid pressure strain terms. The overestimation of velocity and turbulence kinetic energy is due to a linear pressure strain correlation assumption. The overestimated mean axial velocity exhibits an undulated profile at the dilution plane, and the transverse velocity predictions are accurate. The turbulent kinetic energy and shear stress are underpredicted and fail to capture the peak at the center of the plane, as found in the experimental measurement. Nevertheless, the Q-criterion displays that the model has preserved the swirling vortex core till the exit of the combustor, as specified in experimental measurements. The noticed discrepancies are attributed to the linear pressure strain correlation, due to which the model ignores complexities of turbulence, such as non-linear interaction between velocity gradients. Hence, the model cannot accurately capture the complexities of swirling flows, vortex core, and its interaction with the surrounding fluid.}
\end{itemize}
\add{In summary, LPS-RSM generally performed better than the two-equation turbulence models due to anisotropic turbulence consideration. Considering non-linear pressure strain correlation can delineate the drawbacks and improve the accuracy of LPS-RSM for the computation of swirling turbulent flows.}
%	
%--------------------------------------
\section{Concluding remarks}
%--------------------------------------
\noindent
\remove{In this study, isothermal flow in the model can type combustor is simulated. The models assessed are standard $k-\epsilon$, realizable $k-\epsilon$, standard $k-\omega$, SST $k-\omega$ and LPS-RSM model. 
The predictive abilities of models are assessed by comparing the predicted axial velocity, transverse velocity, turbulent kinetic energy and shear stress at the primary hole and dilution hole planes.} 
\add{In this work, the predictive assessability of the turbulence models for isothermal flow in a combustor representing a constituent can combustor of the can-annular configuration used in jet engines has been performed using ANSYS Fluent based on the finite volume method. The RANS-based turbulence models assessed are the two-equation models (standard $k-\epsilon$, realizable $k-\epsilon$, standard $k-\omega$, SST $k-\omega$) and Linear Pressure Strain - Reynolds Stress Model (LPS-RSM). The prediction of the turbulence models for axial velocity, transverse velocity, turbulent kinetic energy, and shear stress have been compared with the experimental data at two different positions (i.e., the primary and dilution hole planes) in the combustor. The two-equation models have generally failed to predict both trends and accuracy of experimental results for the confined swirling flows at both positions, as the model formulations, assuming isotropic turbulence, cannot capture the intricacies of vortex flow and its interaction with the surroundings in confined swirling flows. Amongst the two-equation models, the SST $k-\omega$ model yielded the most accurate, followed by standard $k-\omega$, realizable $k-\epsilon$, and standard $k-\epsilon$ models. Compared with two-equation models, LPS-RSM accounting for the anisotropic turbulence is promising for the confined swirling turbulent flow problems. While the LPS-RSM predictions have consistently captured the qualitative trends per experimental results,  quantitatively, results are overestimated for all flow features, except for the turbulent kinetic energy and shear stress underestimated at the dilution holes plane. These observed discrepancies arise from the linear pressure-strain correlation in the LPS-RSM; this linear assumption is quite simplistic for complex flows. Hence, this study suggests that non-LPS-RSM or more advanced turbulence models can accurately predict the confined swirling flow in combustors.}
%
%%%%%%%%%%%%%%%%%%%%%%%%%%%%%%%%%%%%%%%%%%
\section*{Declaration of Competing Interest}
%%%%%%%%%%%%%%%%%%%%%%%%%%%%%%%%%%%%%%%%%%
\noindent 
% All authors declare that they have no conflict of interest. 
The authors declare that they have no known competing financial interests or personal relationships that could have appeared to influence the work reported in this paper.
%
%The  authors  certify  that  they  have  NO  affiliations  with  or  involvement  in  any  organization or entity with any financial interest (such as honoraria; educational grants; participation in speakers’ bureaus; membership, employment, consultancies, stock ownership, or other equity interest; and expert testimony or patent-licensing arrangements), or non-financial interest (such as personal or professional relationships, affiliations, knowledge or beliefs) in the subject matter or materials discussed in this manuscript.
%%%%%%%%%%%%%%%%%%%%%%%%%%%%%%%%%%%%%%%%%%
%\section*{Acknowledgements}
%\noindent 
%R.P. Bharti would like to acknowledge Science and Engineering Research Board (SERB), Department of Science and Technology (DST), Government of India (GoI) for the providence of the MATRICS grant (File no. MTR/2019/001598). 
%
%--------------- Nomenclature
%\clearpage
\begin{spacing}{1.1}
%\input{Nomenclature.tex}
%\renewcommand{\nompreamble}{\vspace{1em}\fontsize{10}{8pt}\selectfont}
%{\printnomenclature[5em]}
\printnomenclature
\end{spacing}
%
%\clearpage
%\printglossaries 
%
%===========================
%\appendix
%===========================
%
%--------------- Bibliography
%
%\begin{thebibliography}{0000}
%\bibliographystyle{plainnat}
%\bibliographystyle{elsarticle/elsarticle-harv}\biboptions{authoryear}
\noindent 
\bibliography{references}
%
% Bibliographic references with the natbib package:
% Parenthetical: \citep{Bai92} produces (Bailyn 1992).
% Textual: \citet{Bai95} produces Bailyn et al. (1995).
% An affix and part of a reference:
%   \citep[e.g.][Ch. 2]{Bar76}
%   produces (e.g. Barnes et al. 1976, Ch. 2).
% 
% \bibitem[Names(Year)]{label} or \bibitem[Names(Year)Long names]{label}.
% (\harvarditem{Name}{Year}{label} is also supported.)
%\bibitem[]{}
%
%\input{references.tex}
%
%\end{thebibliography}
%
%\clearpage\listoftables
%\clearpage\listoffigures
%\clearpage
%
%\input{tables.tex}
%
%\clearpage
%
%\renewcommand{\thesubfigure}{(\roman{subfigure})}
%
%\input{figures.tex}
%
\end{document}